\documentclass[format=acmsmall, review=false, screen=true]{acmart}

\usepackage{booktabs} % For formal tables
\usepackage[utf8]{inputenc}
\usepackage{graphicx}
\usepackage{array}
\usepackage{color}
\usepackage{textcomp}
\usepackage{caption}
\usepackage{longtable}
\usepackage{amssymb}

\usepackage[ruled]{algorithm2e} % For algorithms

\SetAlFnt{\small}
\SetAlCapFnt{\small}
\SetAlCapNameFnt{\small}
\SetAlCapHSkip{0pt}
\IncMargin{-\parindent}

\usepackage{listings}
\definecolor{grey}{rgb}{0.9,0.9,0.9}
\begin{document}
% Title portion. Note the short title for running heads
\title{Declarative Data Analytics: a Survey}

\author{Nantia Makrynioti}
\affiliation{%
  \institution{Athens University of Economics and Business}
  \streetaddress{76 Patission Str}
  \city{Athens}
  \postcode{10434}
  \country{Greece}}
\email{makriniotik@aueb.gr}
\author{Vasilis Vassalos}
\affiliation{%
  \institution{Athens University of Economics and Business}
  \streetaddress{76 Patission Str}
  \city{Athens}
  \postcode{10434}
  \country{Greece}
}
\email{vassalos@aueb.gr}

\begin{abstract}
The area of declarative data analytics explores the application of the declarative paradigm on data science and machine learning. It proposes declarative languages for expressing data analysis tasks and develops systems which optimize programs written in those languages. The execution engine can be either centralized or distributed, as the declarative paradigm advocates independence from particular physical implementations. The survey explores a wide range of declarative data analysis frameworks by examining both the programming model and the optimization techniques used, in order to provide conclusions on the current state of the art in the area and identify open challenges.
\end{abstract}





\maketitle

\renewcommand{\shortauthors}{N. Makrynioti et al.}

\section{Introduction}
\label{sec: intro}
With the rapid growth of world wide web (WWW) and the development of social networks, the available amount of data has exploded. This availability has encouraged many companies and organizations in recent years to collect and analyse data, in order to extract information and gain valuable knowledge. At the same time hardware cost has decreased, so storage and processing of big data is not prohibitively expensive even for smaller companies. Topic classification, sentiment analysis, spam filtering, fraud and anomaly detection are only a few analytics tasks that have gained considerable popularity in recent years, along with more traditional warehouse queries which gather statistics from data. Hence, making the development of solutions for such tasks less tedious, adds value to the services provided by these companies and organizations, and encourages more people to enhance their work using data.     

Data mining and machine learning are core elements of data analysis tasks. However, developing such algorithms needs not only expertise in software engineering, but also a solid mathematical background in order to interpret correctly and efficiently the mathematical computations into a program. Even when experimenting with black box libraries, evaluating various algorithms for a task and tuning their parameters, in order to produce an effective model, is a time-consuming process. Things become more complicated when we want to leverage parallelization on clusters of independent computers for analysing big data. Details concerning load balancing, scheduling or fault tolerance can be overwhelming even for an experienced software engineer. 

Research on the data management domain recently started tackling the above issues by developing systems that aim at providing high-level primitives for building data science and machine learning tasks and at the same time hide low-level details of the solution or distributed execution. MapReduce \cite{Dean04} and Dryad \cite{Isard07} were the first frameworks that paved the way of large-scale analytics. However, these initial efforts suffered from low usability, as they offered expressive but still low-level languages to program data mining and machine learning algorithms.  Soon the need for higher-level declarative programming languages became apparent. Systems, such as Hive \cite{Thusoo09}, Pig \cite{Olston08} and Scope \cite{Chaiken08}, offer higher-level languages that enable developers to write entire programs or parts of them in declarative style. Then, these programs are automatically translated to MapReduce jobs or Dryad vertices that form a directed acyclic graph (DAG), which is optimized for efficient distributed execution. This paradigm is adopted by other systems, too. Stratosphere \cite{Alexandrov14}, Tupleware \cite{Crotty15} and MLbase \cite{Kraska13} also aim at hiding details of distributed execution, such as load balancing and scheduling, in order to give the impression to the user that she develops code for a single machine.

Apart from the programming model, optimization techniques are another important issue that these systems address. As the declarative paradigm implies that the user expresses the logical structure of a program, there may be many implementations that compute the same result, but differ at efficiency level. Rewritings of the logical structure and physical implementation of a program, which harness the properties of the operators of a language in a way similar to relational query optimization, as well as the use of techniques from the domain of compilers are explored in the context of declarative data science.
  
In this survey, we study systems for declarative large-scale analytics by focusing on two aspects: programming model and optimization techniques. The rest of the paper is organized as follows. Section \ref{sec:scope} describes the scope of the survey, i.e. which data analysis tasks are considered when we present the capabilities of different systems and languages. In section \ref{sec:example} we present the example we will use to analyse the classes of programming models of section \ref{sec:programming_model}, where domain specific languages and libraries are classified into these classes. Section \ref{sec:opt_techniques} examines optimization techniques that are employed by declarative data analysis systems, whereas section \ref{sec:comparison} presents a comparison between the surveyed systems based on specific properties and discusses future directions. Finally, section \ref{sec:conclusion} concludes the survey.

\section{Scope}
\label{sec:scope}
In this work we focus on two popular classes of data analysis tasks: descriptive and predictive analytics.
%Data analysis involves a wide range of algorithms from the areas of data management and machine learning. In this work we focus on two popular classes of tasks: descriptive analytics and predictive analytics.

Descriptive analytics provide insights for the past. This usually involves complex queries on a database system to extract aggregated information, such as sum and average statistics for a collection of records. Data are stored in relations and relational algebra operators are used to form queries on them. On the other hand, predictive analytics study historical data in order to identify trends and produce predictions for future events. Machine learning algorithms for regression, classification and clustering hold a dominant position in this class. In this second category, we will address deterministic machine learning algorithms, either supervised or unsupervised, with the exception of the specific area of deep learning. Many of the algorithms that fall into this category, e.g. linear / logistic regression, support vector machines (SVM) and k-means, are expressed using linear algebra operators, which operate on matrices and vectors, and employ an iterative refinement process to minimize/maximize a given objective function. The aforementioned algorithms are useful for a variety of tasks, including demand forecasting, market segmentation and spam filtering. Probabilistic graphical models, such as Bayesian or Markov networks, and systems such as DeepDive \cite{Sa16}, which allow the definition of inference rules in a declarative manner, as well as statistical relational learning are orthogonal to the scope of this survey and we will not study declarative languages for expressing algorithms in these categories.

Regarding descriptive and predictive analytics, we will examine languages and systems from two perspectives: programming model and optimization techniques. In each of these perspectives we identify specific properties, some of which are also proposed in \cite{Boehm16}, that play important role on the declarative programming paradigm. For example, data abstractions and operators are design choices related to the programming model, whereas efficient plan execution depends on plan optimization techniques. The purpose of the seven properties presented below is to provide a way to measure declarativity under the specific scope of the survey. We discuss them in the following sections, as we describe different categories of data analytics systems.

\begin{itemize}
\item \textbf{Data abstractions}: matrices, vectors, tables and any other data types are exposed as abstractions and are independent from their physical implementation in every system, e.g. distributed or dense/sparse implementations of a matrix. 

\item \textbf{Data processing operators}: basic data processing (DP) operators such as join or group by, which are common in relational databases, need to be supported. 

\item \textbf{Advanced analytics operators}: support for primitives widely used in machine learning (ML), such as linear algebra operators and probability distribution functions.

\item \textbf{Plan Optimization}: users' programs are automatically optimized by the system based on properties and available implementations of the involved operators.

\item \textbf{Lack of control flow}: the user does not have access to control flow constructs, which specify a specific order for the execution of the program.

\item \textbf{Automatic computation of the solution}: the computation of the parameters of a machine learning model needs to be provided by the system in a transparent way to the user. Just as in databases, the algorithmic details of computing the result of a query are not described by the user, in a fully declarative machine learning framework the code that computes the solution should not also be written by the user. A usual example is gradient descent, which needs to be written by the user in many of the systems covered by this survey. Requesting from the user to implement a process that computes the parameters of the machine learning model and as a consequence involves algorithmic details of how the final outcome is produced diverges from the concept of declarative paradigm.

\item \textbf{No need for code with unknown semantics}: operators should not work as second-order functions that take as input user-defined code with unknown semantics.
\end{itemize}

\section{Running example}
\label{sec:example}
To demonstrate the programming model of the various categories of systems, we will use a well-known machine learning algorithm named Linear Regression (LR) supplemented with a filtering process of training data at the beginning. In LR the dependent variable is a real number and is computed as the sum of products between features ($x_i$) and coefficients ($w_i$), as displayed in equation \ref{eq:LR}. Features are independent variables whose values are provided by training data, whereas the values of coefficients are unknown and will be provided by optimizing the Linear Regression model.

\begin{equation}
\label{eq:LR}
\hat{y}=\sum_{i=1}^m{x_i}{w_i},\ m=number\ of\ features
\end{equation}

To find the optimal values of coefficients, LR minimizes / maximizes an objective function. In our examples we minimize Least Squared or Least Absolute Error between predictions and actual values of training observations (equations \ref{eq:L2_LR} and \ref{eq:L1_LR}). The minimization of the objective function is achieved by a mathematical optimization algorithm, such as gradient descent.

\begin{equation}
\label{eq:L2_LR}
\min\sum_{i=1}^n(\hat{y_i}-y_i)^2,\ n=number\ of\ training\ observations
\end{equation}

\begin{equation}
\label{eq:L1_LR}
\min\sum_{i=1}^n|\hat{y_i}-y_i|,\ n=number\ of\ training\ observations
\end{equation}

The particular task at hand is to predict the median value of houses based on a number of features about a suburb, such as crime rate, distance from employment centers, etc. To train the LR model we use the Boston housing dataset \footnote{https://www.cs.toronto.edu/~delve/data/boston/bostonDetail.html}. The dataset contains information about housing in the area of Boston. This information includes statistics about the area, such as the crime rate and nitric oxides concentration, as well as the price of a home. Thus, it can be used to train a model that predicts the median value of a home based on the rest of the attributes. In our example, before training, we preprocess the dataset to filter observations, which are very close to Charles river.

The selected running example covers many of the characteristics typically seen in both supervised and unsupervised machine learning algorithms, as well as functionality that is useful in descriptive analytic queries. In order to implement the example at hand, we first need a suitable set of data structures and operators to express the Linear Regression function and Least Squared Error, which serves as the objective function. In order to minimize/maximize the objective function, we either need an implementation of a mathematical optimization algorithm or iteration constructs to be able to express the iterative process of optimization. Finally, for the preprocessing of training data, we store data in a relation and use a filtering operator, equivalent to the one supported in relational algebra. Although Linear Regression is a supervised ML algorithm, the same building blocks can be used to implement unsupervised ML methods, such as k-means.

\section{Programming Model}
\label{sec:programming_model}
Systems for large-scale data science started as libraries of machine learning / data mining algorithms, but in the more recent years they have evolved into domain-specific languages that support primitives to code data science tasks and emphasize on declarative programming. The latter has become more popular within the last few years based on the argument that the simplification of development of data science tasks offers more flexibility to the users to customize their solutions than black-box libraries targeted to a particular execution engine. A number of approaches is followed to fulfil this goal. In the following subsections we describe each approach in detail and categorize relevant systems according to their design choices on the programming model.

\subsection{Libraries of Algorithms}
\label{subsec:2.1}
In this subsection we present libraries of machine learning / data mining algorithms, such as SVM and k-means, implemented with the primitives of an execution engine. The user is able to write data analysis tasks by calling functions to load data from various data sources, transform data or use machine learning algorithms to analyse them. The programming paradigm is the same as that of a developer writing code for a single machine and calling functions from a third-party library or executing commands from an interactive shell. The difference is that this code / commands will be executed on a specific, possibly distributed, platform. Below we present some examples of such libraries and analyze the programming model using a code snippet \footnote{Similar examples using MLlib can be found on https://spark.apache.org/mllib/} that calls functions from the MLlib \cite{Meng16} library.

MLlib is a scalable machine learning library on top of Spark \cite{Zaharia10}. It includes algorithms for classification, regression, clustering, recommendations and other useful tasks, which the user can either run via interactive shells in Scala and Python or import in her code by calling functions from APIs in both the aforementioned programming languages, as well as Java. For historical reasons, we also mention here the Apache Mahout project \cite{Mahout}, which started as a library of machine learning algorithms implemented on Hadoop. Later the project took a different direction and currently it develops and maintains a Scala DSL (Domain Specific Language) for linear algebra, which we will discuss further in section \ref{sec: ml_systems}. 

Moving away from distributed processing frameworks, MADlib \cite{Hellerstein12} is a library of in-database methods for machine learning and data analysis. The broad know-how that has been developed over the years for database systems makes databases a promising candidate to encompass data analytics. MADlib provides a library of SQL-based machine learning algorithms, which run on database engines. SQL operators are combined with user defined functions (UDF) in Python and C++ that implement iteration, linear algebra and matrix operations, which are prevalent in machine learning algorithms. Popular data analysis tasks, such as classification, clustering and regression, are already implemented in MADlib. Table \ref{tab:1} summarizes the list of algorithms included in each library.

\begin{longtable}{lccc}
\caption{List of algorithms provided by each library} \\
\label{tab:1}       % Give a unique label
%\begin{tabular}{lccc}
\textbf{Algorithm} & \textbf{Mahout} & \textbf{MLlib} & \textbf{MADlib} \\
\hline
\multicolumn{4}{l}{\textbf{Classification}} \\
\hline
Logistic Regression & \checkmark & \checkmark & \checkmark \\
Naive Bayes & \checkmark & \checkmark & \checkmark \\
Complementaty Naive Bayes & \checkmark &  &  \\
Random Forest & \checkmark & \checkmark & \checkmark \\
Hidden Markov Models & \checkmark & 	&  \\
Multilayer Perceptron & \checkmark &  &  \\
Linear SVM &  & \checkmark & \checkmark \\
Decision Trees &  & \checkmark & \checkmark \\
Gradient Boosted Trees &  & \checkmark &  \\
\hline
\multicolumn{4}{l}{\textbf{Regression}} \\
\hline
Linear Regression &  & \checkmark & \checkmark \\
Isotonic Regression &  & \checkmark &  \\
Ordinal Regression &  &  & \checkmark \\
\hline
\multicolumn{4}{l}{\textbf{Clustering}} \\
\hline
k-means & \checkmark & \checkmark & \checkmark \\
Fuzzy k-means & \checkmark &  &  \\
Streaming k-Means & \checkmark & \checkmark &  \\
Spectral Clustering & \checkmark &  &  \\
Canopy Clustering & \checkmark &  &  \\
Bisecting k-Means &  & \checkmark &  \\
Gaussian Mixture	 &  & \checkmark	 &  \\
Power Iteration Clustering (PIC)	& 	& \checkmark	&  \\
\hline
\multicolumn{4}{l}{\textbf{Recommendations}} \\
\hline
User-Based Collaboration Filtering & \checkmark &  &  \\
Item-Based Collaboration Filtering & \checkmark &  &  \\
Matrix Factorization with ALS & \checkmark & \checkmark &  \\
Matrix Factorization with ALS on Implicit Feedback & \checkmark &  &  \\
Weighted Matrix Factorization & \checkmark &  &  \\
SVD++ & \checkmark &  &  \\
Matrix Factorization	 & \checkmark & \checkmark & \checkmark \\
\hline
\multicolumn{4}{l}{\textbf{Topic Modelling}} \\
\hline
LDA & \checkmark	& \checkmark & \checkmark \\
\hline
\multicolumn{4}{l}{\textbf{Dimensionality Reduction}} \\
\hline
SVD & \checkmark & \checkmark &  \\
PCA & \checkmark & \checkmark & \checkmark \\
QR Decomposition & \checkmark &  &  \\	
Lanczos Algorithm & \checkmark &  &  \\	
Principal Component Projection &  &  & \checkmark \\
\hline
\multicolumn{4}{l}{\textbf{Data Mining}} \\
\hline
Apriori & 	&  & \checkmark \\
FP-Growth &  & \checkmark &  \\
Association Rules &  & \checkmark &  \\
\hline
\multicolumn{4}{l}{\textbf{Optimization Methods / Solvers}} \\
\hline
Stohastic Gradient Descent &  & \checkmark &  \\
L-BFGS &  & \checkmark &  \\
Conjucate Gradient &  & \checkmark &  \\
Dense/Sparse Linear Systems &  & \checkmark &  \\
\hline
\multicolumn{4}{l}{\textbf{Time Series Analysis}} \\
\hline
ARIMA &  &  & \checkmark \\
\hline
\multicolumn{4}{l}{\textbf{Statistics}} \\
\hline
Summary Statistics &  & \checkmark & \checkmark \\
Hypothesis Testing &  & \checkmark & \checkmark \\
Streaming Significance Testing &  & \checkmark &  \\
Pearson's Correlation &  &  & \checkmark \\
Cardinality Estimators &  &  & \checkmark \\
\hline
\multicolumn{4}{l}{\textbf{Text Analysis}} \\
\hline
Term Frequency &  &  & \checkmark \\
Sparse TF-IDF Vectors from Text & \checkmark & \checkmark &  \\
Collocations	 & \checkmark & 	 &  \\
Email Archive Parsing & \checkmark &  &  \\
\hline
\multicolumn{4}{l}{\textbf{Utility Functions}} \\
\hline
PMML Export &  & \checkmark & \checkmark \\
XML Parsing & \checkmark &  &  \\
RowSimilarityJob	 & \checkmark &  &  \\
Feature Transformations &  & \checkmark &  \\
Low-Rank Matrix Factorization &  &  & \checkmark \\
\hline
\multicolumn{4}{l}{\textbf{Probabilistic Graphical Models}} \\
\hline
Conditional Random Fields &  &  & \checkmark \\
\hline
%\end{tabular}
%\centering
\end{longtable}

Let us illustrate the programming model of such libraries using the worked example of LR described in section \ref{sec:example}. 

%In the following code snippet we train a Linear Regression model on housing data from Boston suburbs. The task is to predict the median value of houses based on a number of features about a suburb, such as crime rate, distance from employment centers, etc.
 
\begin{lstlisting}[caption={Linear Regression using MLlib library},label={lst:mllib}]
// Features and Labels
JavaRDD<LabeledPoint> trainingData = data.map(line -> {
	String[] tokens = line.split(",");
	String[] features = tokens[1].split(" ");
	double[] feature_vector = new double[features.length];
	for (int i = 0; i < features.length; i++){
		feature_vector[i] = Double.parseDouble(features[i]);
	}
	return new LabeledPoint(Double.parseDouble(tokens[0]), Vectors.dense(feature_vector));
});
        
// Data Filtering 
JavaRDD<LabeledPoint> filteredTrainingData = trainingData.filter(point -> {
	return (point.features().toArray()[3]==0);
});
        
filteredTrainingData.cache();
        
int iterations = 200
double learning_rate = 0.0000001           
        
// Training using Linear Regression  
final LinearRegressionModel model = LinearRegressionWithSGD.train(JavaRDD.toRDD(normalizedTrainingData), iterations, learning_rate);
        
// Generating predictions
JavaRDD<Tuple2<Double, Double>> predictions_labels = normalizedTrainingData.map(point -> {
	double prediction = model.predict(point.features());
	return new Tuple2<Double, Double>(prediction, point.label());
});
\end{lstlisting}
The workflow in listing \ref{lst:mllib} begins with loading the training data from a file and creating a vector of features for each instance. Both the features and the target value are stored in an object that represents a labeled point, i.e. a data point that is annotated with a target value. The set of labelled points are stored in a RDD (Resilient Distributed Dataset) \cite{Zaharia10}, a Spark abstraction for a collection of elements partitioned across the nodes of a cluster. After filtering the points based on a specific feature, a new RDD is created with the same structure, which is used to train a Linear Regression model. The model is then loaded and is capable to give predictions for new data. In this example we output predictions again for training data just to showcase how this can be done using MLlib. 

We can see that the MLlib library provides classes for ML algorithms, such as LinearRegressionWithSGD, and user defined code is needed mainly for preprocessing tasks, like transforming data to the format that is required by algorithm. However, in case users want to use a machine learning algorithm that is not provided by the library, they have to implement it by using the programming model of the underlying platform. 

Because libraries are developed using a specific programming model or language, it depends on those whether the seven declarativity properties described in section \ref{sec:scope} are covered. For example, regarding the property of data processing operators, MLlib is built on Spark, which provides operators for joining or filtering RDDs, whereas in MADlib one can apply relational operators on tables. Also these platforms may or may not support optimization of programs. For these reasons declaritivity properties do not apply in the context of libraries and we will exclude them from the final comparison table at the end of the paper.

\subsection{Hybrids of SQL and MapReduce}
This class includes languages, such as Jaql \cite{Beyer11}, Stratosphere's Meteor \cite{Alexandrov14}, Pig Latin \cite{Olston08} and U-SQL \cite{Usql}, which aim to offer a programming model between SQL and MapReduce. The reasoning behind this approach is based on the fact that SQL is too rigid for some data analysis tasks, but on the other hand MapReduce is too low-level and needs custom code even for the simplest operations. Considerable amount of custom code not only requires effort to be written, but it also increases the time spent on debugging and maintenance. Hence, programs in these languages are sequences of steps as in procedural programming, but with each step performing a single high-level transformation, similar to SQL operations (e.g. filtering and grouping). Meteor also supports domain-specific operations, such as part-of-speech and sentence annotation, duplicate detection, record linkage and other tasks used in data analysis. This way developers can define workflows incrementally by storing intermediate results in variables or using pipe syntax to forward them to the next expression instead of describing the final desired outcome as in SQL. Nevertheless, these intermediate results are generated via declarative operators and the user does not need to describe how these operators are implemented. These languages usually provide a wider range of operators than the limited set of primitives used in MapReduce framework, which leaves a  considerable amount of code to be written by the developer. In case custom code is needed, users can write user-defined functions. 
%An overview of Jaql's, Pig Latin's and Meteor's operators is presented in tables \ref{tab:Jaql},\ref{tab:PigLatin} and \ref{Meteor} respectively. 
Programs written in those languages are translated automatically by the system into lower level code using the programming model of an execution engine, such as MapReduce and Nephele \cite{Battre10}.

Concerning the data model, most of the languages in this class revolve around the concept of tuple known from databases or adopt a JSON-like model. Two important aspects of these models are schemaless records and arbitrary nesting. Schema is not needed or is specified on the fly and fields of a tuple may store non-atomic values, i.e. tuples, bags and maps. The argument behind this feature is that developers of procedural programming are more familiar with storing multiple values in data structures, such as arrays, sets, maps, than with normalizing data to tables. 

Let us now present the aforementioned features with more detail using the implementation of the Linear Regression algorithm in Pig Latin. As stated above, Pig Latin does not provide implementations of machine learning algorithms, so the user has to use its operations and structures to write the code. We use gradient descent for optimizing the weight values and our implementation in Pig Latin also includes code for this procedure. 
%Linear Regression is defined by the product of features and weights, whose values are chosen by an optimization algorithm that minimizes the error between predictions and actual target values. 

In listing \ref{lst:pigLatin} we compute the prediction for each instance, the total error and a single iteration of gradient descent. Given that we have loaded training data from a file to a bag of tuples, i.e. \texttt{input\_data}, we start with filtering housing instances based on their proximity to Charles river. Then, we create a new tuple for each instance consisting of the target value and a nested tuple of features, in order to transform data to a more convenient format for subsequent operations. These preprocessed tuples are stored in bag {\ttfamily input\_data\_preprocessed}. Predictions are generated by computing the product of weights and features for each instance. To combine each feature with its corresponding weight, we implement a UDF. The sum of the differences between predictions and actual target values gives us the total error, which we want to minimize using gradient descent. The derivatives and the weight updates for gradient descent are computed and stored in bags {\ttfamily gradients} and {\ttfamily weight\_updates}. Finally, we store the new weight values in a file. 

\begin{lstlisting}[caption={Linear Regression in Pig Latin}, label={lst:pigLatin}]
-- Filtering observations 
filtered_input_data = FILTER input_data BY f3 == 0;

-- Initialize weights 
weight_data = LOAD '$input_weights' USING PigStorage(',') AS (w0:double, w1:double, w2:double, w3:double, w4:double, w5:double, w6:double, w7:double, w8:double, w9:double, w10:double, w11:double, w12:double);
 
weight_data = FOREACH weight_data GENERATE TOTUPLE($0 .. $12) AS weight_vector;

-- features and labels
input_data_preprocessed = FOREACH filtered_input_data GENERATE $13 AS response, TOTUPLE($0 .. $12) AS feature_vector;

weight_feature_tuples = CROSS weight_data, input_data_preprocessed;
weight_feature_pairs = FOREACH weight_feature_tuples GENERATE response as response:double, WeightFeaturePair(weight_vector, feature_vector) as pairs:bag{t:tuple(weight:double, feature:double, dimension:int)};

-- Generate predictions
predictions = FOREACH weight_feature_pairs {products=foreach pairs generate weight*feature as product:double; prediction = (double)SUM(products.product); GENERATE FLATTEN(pairs) AS (weight, feature, dimension), response as response, prediction as prediction;};

-- Errors between prediction and actual target value
errors = FOREACH predictions {error = (prediction - response); GENERATE weight as weight, feature as feature, dimension as dimension, error as error;};

-- Gradient Descent steps
gradients = FOREACH errors GENERATE weight, dimension, feature*error as feature_error;

weight_updates = FOREACH (GROUP gradients BY (dimension, weight)) {learning_rate=0.0000001/(double)COUNT(gradients); total = SUM(gradients.feature_error); weight_update=learning_rate*total; new_weight=group.weight-weight_update; GENERATE group.dimension as dimension, new_weight as new_weight;};

new_weights = FOREACH (GROUP weight_updates ALL) {in_order = ORDER weight_updates BY dimension ASC; tuple_weights = BagToTuple(in_order.new_weight); GENERATE tuple_weights AS weights;};

final_weights = FOREACH new_weights GENERATE FLATTEN(weights);

STORE final_weights INTO 'pig_test/weight_values' USING PigStorage(',');
\end{lstlisting}

The reason we use UDFs in this code snippet is the lack of support for iteration over columns in Pig Latin.  In our example, we have two vectors, one for features and one for weights, and we need to combine each feature with its corresponding weight. To do so, we need to dive into the contents of the vectors and process them element by element. Although iteration over tuples of a bag is provided by using the {\ttfamily foreach} operator, there is no obvious way to do the same for columns without explicitly stating column numbers or names, which is impossible when we have a large number of columns.It is also important to note that UDFs are treated as a black box and their code is not optimized, as it happens with Pig Latin's operators.

Note that gradient descent is an iterative algorithm. So far we have presented one iteration of it, but in practice we run gradient descent for many iterations or until error converges. However, Pig Latin does not support loop constructs and in order to run this Pig script iteratively, we need an external program, i.e. a driver, in any programming language supporting iteration. In this case we use Python, but any programming language supporting iteration would do. The driver program should initialize weights with zero and run the Pig script for a number of times passing as parameters the files for feature and weight datasets. After each round, the file storing weight values is deleted and created again with the new weights that are computed by gradient descent. It is clear that this approach is far from efficient. Due to the lack of iteration support we are forced to write and read data from files, and repeat preprocessing steps, which would be avoided if we were able to use variables for intermediate results after each iteration.

%\begin{lstlisting}[caption={Driver program to run Pig script of listing \ref{lst:pigLatin} iteratively}, label={lst:driverPig}]
%#!/usr/bin/env python
%import sys, random, os
%
%num_features = sys.argv[1]
%command='hadoop fs -rm pig_test/weight_values.csv'
%os.system(command)
%weights = []
%for i in range(0, int(num_features)):
%	weights.append(str(0))
%
%input_weights = "weight_values.csv"
%weight_file = open(input_weights, 'w')
%weight_file.write(",".join(weights) + "\n")
%weight_file.close()
%command='hadoop fs -put weight_values.csv pig_test/'
%os.system(command)
%
%for i in range(1, 200):
%	input_data = 'pig_test/housing_data_features.csv'
%	input_weights = 'pig_test/weight_values.csv'
%	command='hadoop fs -rm -r pig_test/weight_values'
%	os.system(command)
%	command = 'pig -f linear_regression.pig -p input_data=\'pig_test/features.csv\' -p input_weights=\'pig_test/weight_values.csv\'
%	os.system(command)
%	command='hadoop fs -rm pig_test/weight_values.csv'
%	os.system(command)
%	command = 'hadoop fs -mv pig_test/weight_values/part-r-00000 pig_test/' 
%	os.system(command)
%	command = 'hadoop fs -mv pig_test/part-r-00000 pig_test/weight_values.csv' 
%	os.system(command)
%\end{lstlisting}

Regarding the data model, we can observe that during the loading of the input data a schema is defined. However, this is not necessary and the LOAD command could have ended after the delimiter definition. Nested data structures are also used since the initial bag of tuples that stores scalar values is transformed to the bag called {\ttfamily input\_data\_preprocessed}, where each tuple consists of a scalar and a nested tuple.

Overall, two design choices characterize this programming model: the mix of procedural and declarative programming and the development of flexible data models to allow for semi-structured and unstructured data. However, the main limitation of these languages is lack of iteration support. There are some operators for iteration over data, such as the FOREACH operator of Pig Latin, but the user is not able to declare that a given group of operators will be repeated for a number of times or as long as a certain condition holds. Iterative processes are common in machine learning algorithms, which are frequently used for more advanced data analysis tasks. Given the lack of control flow and the lack of automatic solving methods to provide the values of the model parameters, the approach of this class of systems has been abandoned in the context of machine learning, as different directions that we will discuss in the rest of the survey proposed more efficient architectures for the described limitations. 

Hence, this class of systems supports five out of the seven declarativity properties in the context of data analytics, namely data abstractions, data processing operators, plan optimization, which will be discussed in section \ref{sec:opt_techniques}, lack of control flow and UDF free operators.
%Nowadays, systems like Pig Latin are mainly used to execute SQL-like queries or preprocessing code on top of MapReduce clusters before feeding the data to a machine learning algorithm.

\subsection{Extensions of MapReduce}
In this section, programming models propose extensions to the MapReduce model. They employ the idea of passing first-order functions to operators that are integrated to an imperative/functional language, and extend it mainly at two directions. First, the set of operators provided by these models is enriched compared to MapReduce, which is limited to only two operators. For example, join, filter, union or variants of the map operator, such as flatmap, are implemented by these models, which try to offer an out of the box solution instead of the tedious fitting of these common operations to the MapReduce model. For example, Spark integrates such operators to Scala, Java and Python, DryadLINQ \cite{Yu08} to LINQ, which is a set of constructs operating on datasets and are enabled in .NET languages, whereas Tupleware uses Python. The second important aspect is that these programming models allow for arbitrary dataflows, which combine operators in any order and each logical plan can form a directed acyclic graph (DAG) instead of a sequence consisting of a single map and reduce function.

Programming models of this category also take iterative processes into consideration by allowing in memory processing. This enables the development of machine learning algorithms, whose optimization methods are mainly iterative processes. However, when it comes to distributed processing, supporting iteration can be quite challenging, as updates and propagation of global variables to every node of the cluster at the end of each iteration are not trivial. A few approaches are followed by the programming models of this category to address these issues. DryadLINQ has very limited support of shared state, as it allows read-only shared objects and computation results become undefined if any of them is modified. This is an important restriction of the DryadLINQ model with regard to machine learning algorithms, as they usually optimize a model by updating global parameters after each iteration. Spark and Tupleware use the concepts of Accumulator and Context respectively. Both objects can be updated only by associative and commutative operations. Spark also provides another type of shared variables, which are called Broadcast variables and are analogous to the read-only shared variables supported by DryadLINQ. Flink \cite{Flink} (initially known as Stratosphere) has also Accumulators, which share similar features with those in Spark, but their partial results are only merged at the end of a Flink job. For computing simple statistics among iterations, Flink provides Aggregators, which can be used to check for convergence after each iteration step.

Although the design of MapReduce is not well-suited for iterative workloads \cite{Doulkeridis14}, there has been effort on extending the original programming model to support loop constructs. One of these efforts is HaLoop \cite{Bu10}, which provides functions to specify a termination condition and the input data of a loop, as well as mechanisms for caching loop-invariant data. It also schedules map and reduce tasks that occur in different iterations but access the same data on to run on the same machine. Another system that implemented a set of features for efficient handling of iterative processes in MapReduce is Twister \cite {Ekanayake10}. The features include long-running map/reduce tasks that read loop-invariant data only once, programming extensions that define broadcast variables or combine all reduce outputs to a collective output, and a streaming-based runtime so that intermediate results between iterations do not need to be written on disk, but are disseminated from their producing to their consuming map/reduce tasks. However, due to the wide adoption of Spark which generalized in-memory processing for DAGs, these solutions did not prevail.

Regarding the data model, these systems represent data as immutable collections of records, which can be distributed and individually processed by a machine of the cluster. Data types of the elements in a record are based on the data types provided by the host language. Collections are also represented by objects of the host language. For example DryadLINQ datasets are managed via DryadTable objects in .NET, whereas Spark RDDs (Resilient Distributed Datasets) are Scala objects. We use the following implementation of our running example in Spark to showcase the features described above.

In the code of listing \ref{lst:spark}, training data are represented using a RDD of objects of class {\ttfamily LabeledPoint}. Each of these objects consists of a real target value and a vector of features. First, we filter training observations based on the value of the third feature in each vector. Weights are also stored in a vector, which is transformed to a broadcast variable (read-only variable) in order to be available to every node of the cluster. Using a map operator, we then compute the error between the prediction and the target value and store it as a RDD of double values. A reduce function over the RDD of errors returns the total error. The gradient descent algorithm is also implemented in a similar manner. A map function computes the partial derivative of each error value, whereas the gradient is given by aggregating over all partial derivatives.

\begin{lstlisting}[caption={Linear Regression in Spark}, label={lst:spark}]

import breeze.linalg.{DenseVector => BDV}
import breeze.linalg.{DenseMatrix => BDM}

def linearRegression (data: RDD[LabeledPoint], 	sc: SparkContext)  {

//Labels and Features    
val trainingData = data.map { line =>
      
	val parts = line.split(',')
     val feature_vector = new Array[Double](parts.length-1)
     for(i <- 0 to (parts.length -2)){
     	feature_vector(i)=parts(i).toDouble
     }
     LabeledPoint(parts(parts.length-1).toDouble, Vectors.dense(feature_vector))
      
}

//Filtering observations 
val filteredTrainingData = trainingData.filter {point =>
	val features = point.features.toArray
	features(3) == 0
}

filteredTrainingData.cache();
    
//Weight initialization
val weights = Vectors.zeros(13)
    
val features = parsedData.map { point =>
      
	point.features
      
}.cache()
    
//Number of training examples
val numInstances = sc.broadcast(features.count())
    
//Use of var to define a mutable reference to weights, as they are
//reassigned after each iteration of gradient descent
var broadcastWeights = sc.broadcast(BDV(weights.toArray))
   
//Gradient descent
for(i <- 1 to 200){ 
    
	val errors = parsedData.map { point => 
          
	val features = BDV(point.features.toArray)
	val features_transpose = features.t
	val label = point.label
          
	(label - (features dot broadcastWeights.value)) * (label - (features dot broadcastWeights.value))
          
}
     
      
val totalError = errors.reduce((a, b) => a+b)
val totalError_m = totalError/(numInstances.value)
println("TotalError: " + totalError_m)
        
val newWeights = computeGradients(parsedData, broadcastWeights, numInstances)
        
broadcastWeights = sc.broadcast(BDV(newWeights.toArray))

  
def computeGradients (data: RDD[LabeledPoint], 	inputWeights:org.apache.spark.broadcast.Broadcast[breeze.linalg.DenseVector[Double]], numInstances:org.apache.spark.broadcast.Broadcast[Long]) : breeze.linalg.DenseVector[Double] = {
    
	val learning_rate = 0.0000001
    
	val gradients = data.map { point => 
          
		val features = BDV(point.features.toArray)
		val features_transpose = features.t
		val label = point.label
          
		-(2.0/numInstances.value)*(features * (label - (features dot inputWeights.value)))                    
	}
    
        
	val totalGradient = gradients.reduce {case (a:(breeze.linalg.DenseVector[Double]), b:(breeze.linalg.DenseVector[Double])) => 
		
		BDV((a.toArray, b.toArray).zipped.map(_ + _))}
      	
     val weights = (inputWeights.value - (learning_rate*totalGradient)).toDenseVector
	return(weights)
}
  
\end{lstlisting}  

Spark does not have complete support for linear algebra. Though it provides data structures for vectors and matrices, as well as an API of basic linear algebra operations, the design of those expose properties of distribution to the user by giving her the possibility to choose between local or distributed implementations, as well as different formats. For example, distributed matrices are supported in four different formats, RowMatrix, IndexedRowMatrix, CoordinateMatrix and BlockMatrix. The details of these formats may not be intuitive to machine learning experts, who just need a logical abstraction of a matrix / vector to express linear algebra computations in their algorithms. Another important point is that the provided API does not include the same functionality among the different implementations. For example, one can use the {\ttfamily transpose} function to get the transpose of a local matrix, but the corresponding function is not available for RowMatrix format. Such differences may make users to choose types based not only on format characteristics, but also on the available functionality. For these reasons, we use the Breeze numerical processing library \footnote{https://github.com/scalanlp/breeze} for some linear algebra operations involved in the running example, whereas iteration and aggregation are driven through map and reduce operators of Spark. Unfortunately, this means that code is needed to convert data from the RDD format to Breeze data structures, which is exactly what \texttt{BDV} and \texttt{BDM} functions in listing \ref{lst:spark} do.

At the declarativity front, systems in this category support three out of seven properties. Data processing and at least a subset of linear algebra operators is supported. DAGs of operators are also optimized based on a number of techniques. Regarding the rest of the properties, independence of data abstractions is not fully provided by the systems discussed in this category, as the user has access to caching and distribution properties of the data. For example, in listing \ref{lst:spark} we use the \texttt{cache} function to cache the filtered training observations and the \texttt{broadcast} function a bit further down the road to define that the weights vector is going to be broadcasted to every node. Hence, the user is required to have an understanding about the physical implementation of the data abstraction used. Moreover, because in these systems operators are essentially second-order functions, code that is given as argument to the functions is written in existing imperative or functional languages, which makes it quite challenging to optimize. In the following sections, we describe how some systems, such as Flink and Tupleware attempt to overcome this problem by doing static analysis of the code and applying optimizations from the domain of compilers.  Also the integration of operators into imperative or functional languages exposes control flow constructs to the user.  Finally, the algorithm that computes the parameters of a machine learning model is coded by the user and no out of the box solvers, like gradient descent, are provided.

\subsection{Systems Targeted to Machine Learning}
\label{sec: ml_systems}
\subsubsection{Machine Learning Frameworks}
\label{sec:ml_frameworks}
Machine learning algorithms usually involve linear algebra operations, probability distributions and computation of derivatives. Based on these characteristics, systems in this category provide domain specific languages (DSLs) and APIs that support data structures for matrices and vectors, as well as operations on them, distribution and deep learning functions, and automatic differentation. Some representative systems of this class are SystemML \cite{Ghoting11}, Mahout Samsara \cite{Mahout}, TensorFlow \cite{Abadi16}, BUDS \cite{Gao17} and MLI API \cite{Sparks13}. Not every one of them supports all of the aforementioned features and degree of declarativity also differs. We emphasize such differences using the example of linear regression written in DML, the declarative language used in SystemML.

In listing \ref{lst:systemml}, almost every computation is expressed as a linear algebra operation between matrices and vectors. SystemML has more recently added frames, a data structure for tabular data with limited support for transformations, which however does not include filtering of rows. Other systems in this category does not provide any structure for relational or tabular data. Due to this, we omit filtering of training observations in example \ref{lst:systemml} and load data directly from a .csv file to a 506x14 matrix. We also create a one-column matrix for storing weights and initialize it with zeros. To implement gradient descent we use a loop, where we compute predictions and gradients, as well as update {\ttfamily weights} and {\ttfamily total\_error} in each iteration.

\begin{lstlisting}[caption={Linear Regression in DML, one of the languages supported by SystemML}, label={lst:systemml}]
# Read .csv file
data = read("/home/nantiamakrynioti/Downloads/housing.csv",sep=",", rows=506, cols=14,format="csv");

features = data[,1:(ncol(data)-1)];

# Labels of training set 
labels=data[,ncol(data)];

# Number of training examples
m = nrow(labels);

# Choose a learning rate
learning_rate=0.0000001;
# Number of iterations
iterations = 200;

# Initialize weights to zero
weights = matrix(0,rows = ncol(features),cols=1);

total_error = matrix(0,rows = iterations,cols=1);

# Gradient descent
predictions = features%*%weights;
error = predictions - labels;
for( i in 1:iterations){
	gradients = (t(features)%*%error)/m;
	weights = weights - learning_rate*rowSums(gradients);
	predictions = features%*%weights;
	error = predictions - labels;
	total_error[i,1] = t(error)%*%error;
	total_error[i,1] = total_error[i,1]/(2*m); 
}

error = colSums(features%*%weights - labels)/nrow(features);
\end{lstlisting}
Iteration is supported via the usual constructs ``for" and ``while", which make it possible to express the entire algorithm in DML, with no need for driver programs in other languages. In addition to this, linear algebra operators and matrices provide a more intuitive way to express computations and keep the code succinct. Similar code could be written in the rest of the systems in this category but with a few differences hidden in the details. 

Mahout Samsara, which is a Scala DSL (Domain Specific Language) for linear algebra baring a R-like look and feel, allows users to decide about specific representations of data abstractions and data flow properties. For example, the user can choose between local and distributed or dense and sparse matrices, as well as specify caching and partitioning of the data. These options contradict to some extent the notion of declarativity, which implies that data abstractions should be independent of physical implementations and data flow properties should not be exposed to the user. Tensorflow's API in Python and other languages provide the expressivity of SystemML, but the user can also leverage automatic generation of gradients and avoid writing code for this computation. TensorFlow also focuses on the training of neural networks, as it supports built-in functions commonly used in deep learning, such as SoftMax, Sigmoid and ReLU. SystemML recently added a small set of functions for deep learning, but TensorFlow is a lot more advanced in this area. Regarding declarativity, Tensorflow also exposes details of physical data representation to the user with sparse and dense tensors. Another API, MLI API, is implemented on Spark and apart from linear algebra operators, it also supports relational operators. This is an interesting approach as the API puts data processing and machine learning under the same umbrella. However, the project is no longer under active development and the authors claim that many of these ideas have been integrated into Spark MLlib and Keystone ML \cite{Sparks17}. Last but not least, BUDS is fundamentally declarative in the sense that each computation is a list of dependencies among variables and variable updates happen recursively according to those dependencies. There are no control flow constructs and iteration is expressed via recursion. The ``for" construct serves for parallel execution of variable expressions, instead of typical looping.

Hence, there is a twofold situation in this class of systems as far as declarativity properties are concerned. Although the main data abstraction used are matrices and vectors, some of the abstractions are not independent from their physical implementation. Two properties that are common pretty much across the class are the lack of data processing operators and the support of more advanced analytics operators, such as linear algebra and deep learning functions. Also at least some basic plan optimization happens in most of the frameworks. As we will describe in Section \ref{sec:opt_techniques}, some of the languages presented here are accompanied by optimizers developed specifically for them, while others are translated to existing languages and rely on their optimizers or on optimizations performed by the execution engine. Regarding iterative processes, we see that some systems provide loop constructs, while others, e.g. TensorFlow, support automatic differentiation and mathematical solvers for encapsulating the training. Finally, the need for user defined code outside the set of supported operators is usually limited in this category.

\subsubsection{Model Selection Management Systems}
\label{sec:msms}
At the machine learning front, there are a few systems that aim at defining tasks instead of algorithms. These systems attempt  to be accessible to non-experts and provide a very high-level declarative language above a library of algorithms. The user can define the type of task, such as classification, regression, clustering etc, as well as the data to be used with this high-level language. A model selection management system tries different featurization techniques and machine learning algorithms from its library on the data provided by the user, tunes the parameters of these algorithms and determines an effective combination based on quality and time performance. A trained model using the selected algorithm and parameters is returned to the user, along with a summary of its accuracy.

This idea was presented in a few systems and papers, such as MLbase \cite{Kraska13}, AutoWeka \cite{Thornton13}, AzureML \cite{Azure} and \cite{Kumar2016}. The selected algorithm and parameters may not be optimal, but systems in this category intend to guarantee reasonable trade-offs between efficiency and quality for many scenarios, just like a DBMS system. They hide the complexities of choosing an effective algorithm, searching the parameter space and evaluating the accuracy of a model,  in order to minimize the expert knowledge the user needs to have. While there are differences in certain aspects of model selection management systems, for example executing code on a single node versus distributed processing or the use of high-level languages versus graphical user interfaces, the general intention of these systems is to help improve the iterative process of model selection.

This category of systems does not conform to the declarativity properties we examine in this survey. That is because the definition of machine learning tasks may be done via a graphical user interface. It is not necessary to provide a language for this purpose. Even when a language is provided, it consists of a limited number of very high-level functions, e.g. for loading data or starting a classification task, and does not involve the types of lower-level abstractions, constructs and operators we examine in the context of declarativity.

\subsection{In-database machine learning}
In-database machine learning regards approaches that attempt to model machine learning algorithms inside a database. This is not the same with providing a library of algorithms as UDFs for the user. Approaches in this class follow three directions: array databases, mathematical optimization on relational data and extension of the SQL language with linear algebra data types and operations.

\subsubsection{Array databases}
Scientific data, such as astronomical images or DNA sequences, are commonly represented as arrays. Array databases propose a data model that fits better ordered collections of data than the relational model. In addition to this, arrays can be used to represent matrices and vectors widely involved in machine learning algorithms.

A prevalent system in this category, SciDB \cite{Stonebraker11} uses the following data model: an array that consists of dimensions and attributes. Dimensions serve as indices / coordinates of each cell, whereas attributes are the actual contents of a cell. Cells may either contain tuples of attributes or may be empty (null), so that both dense and sparse arrays can be represented. Each attribute is of a primitive type (int, float, char). SciDB also supports the definition of complex types by the user in a similar way as Postgres.

SciDB offers a functional language and a query language with similar syntax to SQL. Since the data model is based on arrays, operators provide common operations on array data. Those include slicing an array along a dimension, subsampling a region of an array, filtering attribute values in an array, applying a computation on the cells of an array and combining cells from two arrays. Basic linear algebra operators, such as matrix multiplication and transpose, are also supported. The implementation of Linear Regression using SciDB in listing \ref{lst:scidb} gives more details about its programming model. 

\begin{lstlisting}[caption={Linear Regression in SciDB}, label={lst:scidb}]

-- Loading feature values and initial weight values to arrays
create array features <instance_id:int64, feature_id:int64, val:double>[i=0:6577];

load(features, '/home/hduser/features.csv',-2,'CSV');

store(redimension(features, <val:double>[instance_id=0:505:0:1000; feature_id=0:12:0:1000]), features_2d);

create array weights <feature_id:int64, val:double>[i=0:12];

load(weights, '/home/hduser/weights.csv',-2,'CSV');

store(redimension(weights, <val:double>[feature_id=0:12:0:1000, j=0:0:0:1000]), weights_2d);

-- Multiplying feature and weight arrays to produce predictions
store(build(<val:double>[row=0:505:0:1000; col=0:0:0:1000],0),C1);

store(gemm(features_2d, weights_2d,C1), predictions);

create array labels <instance_id:int64, val:double>[i=0:505];

load(labels, '/home/hduser/labels.csv',-2,'CSV');

store(redimension(labels, <val:double>[instance_id=0:505:0:32, j=0:0]), labels_2d);

-- Compute errors between predictions and actual values
store(project(apply(join(predictions,labels_2d), e, predictions.gemm - labels_2d.val), e), errors);

store(build(<val:double>[row=0:12; col=0:0],0), C);

-- Compute gradients
store(gemm(transpose(features_2d), errors ,C), gradients);

store(apply(gradients, d, gradients.gemm/506), gradients_2);

aggregate(gradients_2,sum(d),col);

-- Compute new values for weights
store(apply(cross_join(weights_2d, aggregate(gradients_2, avg(d) as m)), val2, weights_2d.val-0.00001*m), new_weights);

store(redimension(new_weights, <val2:double>[feature_id=0:12]), new_weights_2);
\end{lstlisting} 
The first query creates a one-dimensional array, where each cell is a tuple consisting of a feature id and a feature value. Feature values in a csv file are loaded to this array using the {\ttfamily load} command. We then move on to the implementation of the Linear Regression model, without filtering of the training observations. That is because the array data model of SciDB does not support filtering of rows according to a boolean expression. This is justified by the fact that SciDB targets ordered data, such as images, where rows are related to each other. Filtering of columns or slicing a subset of the array based on pairs of coordinates is supported, though. So relational data processing may not be supported, but other data processing that are applicable on the array data model are provided.

Back to the code in listing \ref{lst:scidb}, in order to be able to use linear algebra operators for matrix multiplication or transpose, we need to transform the initial array to a new one that is two-dimensional and where each feature value is stored in a different cell as a tuple with a single attribute and feature id has become the second dimension. The same process is also followed for creating arrays for weight values and labels. When creating an array, the user can also specify chunk size, which determines the maximum number of cells in each chunk. Although that does not seem necessary for all cases as SciDB can automatically choose a chunk length based on schema, in the presented implementation of Linear Regression {\ttfamily gemm} operator complained about the chosen chunk size being small. This type of properties does not concern the logical structure of a data analysis problem and tend to mix logical with physical implementation, which contradicts the purpose of declarative programming. 

The rest of the queries implement the steps of a gradient descent iteration, starting with multiplying weight and feature matrices to compute predictions using SciDB's function {\ttfamily gemm}. Despite matrix multiplication and transpose are supported by specific operators / functions, element-wise addition / subtraction / multiplication are implemented by joining tables and applying a transformation on their cells as it is shown in the creation of array {\ttfamily errors}. So, regarding linear algebra the design of SciDB follows two strategies: complicated linear algebra operators are provided by the database, whereas simpler ones are implemented / emulated using array-based operators. Code below implements a single iteration of gradient descent, as there are no constructs to express iterative processes in SciDB. Similarly to Pig Latin, SciDB operators can be embedded to an imperative language or use a driver program to execute SciDB queries multiple times, but of course this is not optimal performance-wise and more importantly it mixes declarative with imperative programming. 

%
%
%
%For example, creating a 10x10 array with each cell storing a 2-attribute tuple in SciDB's SQL-like query language is done as follows:
%
%\begin{lstlisting}[caption={Creating an array in SciDB}]
%CREATE ARRAY A <att1: int, att2: float> [I=1:10, J=1:10]
%\end{lstlisting}
%
%Taking a projection of the 5th row of array $A$ is as simple as the following query:
%
%\begin{lstlisting}[caption={Getting slice of an array in SciDB}]
%Slice ( A, I = 5);
%\end{lstlisting}

TileDB \cite{Papadopoulos16} is another system relevant to array data management. It supports the same data model as SciDB, but is currently a storage manager for array data rather than an array database. For reasons of completeness, we give a brief description of its capabilities. The storage manager module is accessed via a low level C API, which includes functions for initializing and finalizing an array (freeing its memory), loading and retrieving the schema of an array, reading and writing to an array, iterating over its elements, as well as synchronizing between files and merging of array updates. As TileDB is not an array database, for now it does not include query optimization. The API serves as an interface between the user and the storage manager, but users' programs are not transformed to logical / physical plans and are not optimized by the system.

Array databases have been successfully used for data science tasks on scientific data, as those are naturally represented with arrays. However, iterative processes in machine learning, such as gradient descent, cannot be natively supported using the primitives of these languages. Although workarounds in other languages can be developed for this and an extension to the original model \cite{Soroush15} has been proposed, iteration is still not integrated into the SciDB query language and as a result optimized by a query optimizer. Systems in this category support all declarativity properties, apart from automatic computation of the solution. Their data model and processing operators operate on arrays. They support at least a subset of linear algebra operators and their SQL-like query language lacks control flow structures. Semantics for both types of operators are well-defined and do not depend on user defined functions. Finally, despite not as advanced as in relational databases, some query optimization techniques are implemented in SciDB.

For relational data to be ported to the array data model, denormalization is necessary, which may result in losing important information that is embedded in the relational structure. In the next sections, we analyse the other two approaches for in-database machine learning that aim to address this issue and perform data science directly on relational data. 

\subsubsection{Mathematical Optimization on Relational Data}
In-database mathematical optimization regards the expression of mathematical optimization problems on relational data. Relational databases are extensively used. So the ability to integrate the expression of mathematical optimization problems with a database offers great value, as it would eliminate the need to manually export and import data to different systems. As many machine learning algorithms are deep down numerical optimization problems, modeling such problems inside a database would also allow modeling machine learning tasks.

One such system that allows modeling mathematical optimization problems on relational data is the LogicBlox database \cite{Aref15}, \cite{Sanchez}. The LogicBlox database supports the expression of linear, quadratic and SAT programs using LogiQL, an extended variant of Datalog. The user needs only to define the objective function, the constraints and any other business logic in LogiQL. Then, a rewriting process, which is called grounding, transforms the relational form of the convex optimization program to a matrix format, that is consumed by an external solver. For example, in case of linear programming (LP), grounding involves the automatic creation of predicates that represent the LP instance in its canonical form, i.e. the matrix A and vectors c and b in $max \,  \{c^Tx | x{\geq}0, Ax{\leq}b\}$, as well as the rules that populate these predicates during runtime. As soon as the data are marshaled to the data structures supported by the solver, the optimization process of the problem begins. The solver responds with the computed solution, which is finally stored back in the database and can be accessed via typical LogiQL queries and commands.

The LogiQL code in listing \ref{lst:logiql} implements Linear Regression as a linear program \cite{Makrynioti18}.  We assume that the type and arity of each predicate are defined earlier in the program and that the extensional database (EDB) predicates, i.e. \texttt{observation}, \texttt{feature\_value} and \texttt{target}, are populated by user's data. So the code snippet begins with defining the rules that populate the intensional database (IDB), i.e. the predicates that are not explicitly imported by the user. The predicate {\ttfamily totalError} is the objective function, which is denoted using the annotation \texttt{lang:solver:minimal(totalError)}. The total error is the sum of the individual absolute errors between the target value and the prediction of an observation. The absolute error is expressed using two linear constraints, which state that the absolute error is greater or equal than both the value of the error and the negative value of the error. The variables of the linear program, i.e. the ones whose values will be given by the solver, are defined using the annotation {\ttfamily lang:solver:variable}.

\begin{lstlisting}[caption={Linear Regression in the LogicBlox database using Datalog}, label={lst:logiql}]						
//Weight is a LP variable
lang:solver:variable(`weight).
		
//Abserror is a LP variable		
lang:solver:variable(`abserror).

observations(i)

//Data filtering
filtered_observation(i) <- observation(i), f="CHAR", feature_value[f, i]=v, v=0.0f.

//IDB rule to generate predictions		
prediction[i] = v <- agg <<v=total(z)>> filtered_observation(i), feature_value[f, i]=v1, weight[f]=v2, z=v1*v2.
		
//IDB rule to generate error between predicted and actual target values
error[i] = z <- prediction[i] = v1, target_actual[i] = v2, z = v1-v2.

//IDB rule to generate the sum of all errors
totalError[] = v <- agg << v=total(z) >>  abserror[i]=z1, z=z1.
lang:solver:minimal(`totalError).

//Constraints to implement that abserror is the absolute value of error
filtered_observation(i), abserror[i]=v1, error[i]=v2 -> v1>=v2.
filtered_observation(i), abserror[i]=v1, error[i]=v2, w=0.0f-v2 -> v1>=w.
\end{lstlisting} 

MLog \cite{Li17} has a similar design to LogicBlox's framework. It offers a declarative language, whose data model is based on tensors. The user can create a tensor using a CREATE command, similar to SQL, and manipulate the tensor with a set of operations, including slicing and linear algebra operations. Each MLog program consists of a set of rules, which create TensorViews based on expressions involving operations over tensors. The head of the rule is the name of the TensorView, whereas the body of the rule consists of operations between tensors. The concept is similar to relational views, which are the result of a query or intensional database rules in LogiQL / Datalog. The user can also run MLog queries, which correspond to mathematical optimization problems. These queries find the optimal instances for a number of tensors that minimize or maximize a TensorView defined earlier in the program. This is similar to the LogiQL annotation "lang:solver:minimal" that accompanies the predicate of the objective function. The solution to the MLog queries is computed and provided by TensorFlow, which is also an external system to the database. MLog programs are automatically translated to Datalog programs by representing tensors as a special relation type. Based on this representation the authors claim that the MLog language can potentially be integrated with SQL, although they have not implemented this yet. Finally, Datalog programs are subsequently translated to TensorFlow code.

%MLog programs are automatically translated to Datalog programs by the system, in order to leverage data dependencies and standard database statistics to provide more optimization opportunities of the initial code. Despite offering a high level language that allows easier and more succinct programs for machine learning, program optimization is eventually performed on Datalog code. MLog's data model is mapped to the relational model by representing a tensor as a special relation type. Finally, Datalog programs are translated to TensorFlow code, which is also an external system to the database and this is where all the solution of the mathematical optimization problem actually takes place.  

%Other relevant to this category systems are ReLooP \cite{Mladenov16} and Saul \cite{Kordjamshidi15}, which allow declarative specification of mathematical optimization problems in imperative languages, such as Python or Scala. These are implemented by embedding a domain specific language into an imperative language that allows declarative expression of the main components of an optimization problem, i.e. the objective function, the constraints and the variables. Relational data can be queried via drivers and APIs written in each imperative language. However, the relational database is not aware of the mathematical optimization process and the two systems remain separate from each other. As a consequence, we do not consider these works in-database solutions for mathematical optimization.

To summarize, systems in this category follow the model+solver paradigm, where the user defines the logical structure of the model, whose solution is delegated to a black box. For the declarative definition of the model, either a set of linear algebra operators are provided or relational operators are used to implement equivalent results, wherever it is possible. Thus, support for iteration is not that necessary using this approach, as the computation of the solution which is usually iterative is not implemented by the user. Data processing tasks, which are useful for preprocessing, feature engineering and other types of data analytics, are also easily expressed in these systems using relational algebra.

Systems in this class cover six out of seven declarativity properties. Depending on the data abstraction that they use, being relations or tensors, they support natively either linear or relational algebra operators. Plan optimization is implemented via standard techniques for SQL or Datalog queries in relational databases. Moreover, because the supported languages are purely declarative, the user does not have control over the execution order of the commands inside a program.  Operators do not serve as second-order functions, though code in imperative languages, whose semantics is unknown and cannot be optimized by the database, is supported via UDFs. Finally, the computation of the model parameters is not implemented by the user, rather it is provided by the system. 

\subsubsection{In-database Linear Algebra}
\label{sec: indb_lr}
Following the same direction of modelling machine learning algorithms inside the database, a very recent approach \cite{Luo17} demonstrates the support for linear algebra inside relational databases by extending the relational model with three data types, matrix, vector and labelled scalar, as well as a set of operations over the aforementioned types. These types can be used for attributes when creating a table, i.e. a column in a table can store a matrix or a vector. Moreover, within this approach, common linear algebra operations are fully integrated into the SQL language. The user can perform matrix-matrix or matrix-vector multiplications as part of a query or use standard SQL aggregations on the elements of a vector. The new set of operations also includes aggregate functions for composing instances of the new data types, for example the function VECTORIZE creates a vector from a collection of labelled scalars and ROWMATRIX / COLMATRIX create a matrix from a set of vectors.

Both this approach and the systems for mathematical optimization in previous section share similar benefits. Moreover, they could also be combined and instead of having a new declarative language based on tensors or emulating linear algebra operators using relational operators, one could implement the described extensions in the relational model, but still use a black-box solver to provide the solution.

This approach also scores high in declarativity as it extends an already declarative language, i.e. SQL, to support mode advanced analytics operators from linear algebra. The only property that is not covered is the computation of the model parameters, which still needs to be written by the user in terms of the extended version of SQL and whose iterative nature is not well-supported by this type of languages.

\subsection{Declarativity and Calculus}
\label{sec: declaritivity_calculus}
In previous sections, we showcased through the example of linear regression that users often need to implement challenging mechanics of machine learning algorithms. A common one is the definition of gradients. In all of the systems we surveyed except from TensorFlow, the user needs to define the gradients of the objective function manually, which is time consuming and requires a certain familiarity with differentiation in mathematics. A key component to automate this part that is currently missing from systems for declarative machine learning is either symbolic or automatic differentiation \cite{Baydin17}. Implementations of symbolic differentiation in computing environments like MATLAB \cite{MATLAB} and Mathematica \cite{Mathematica}, and automatic differentiation in TensorFlow and a few frameworks for deep learning indicate the usefulness of this feature in the machine learning toolkit.

Given a function by the user, symbolic differentiation can generate the derivative for every math expression involved and combine them using rules, such as the chain or the sum rule, to compute the final derivative. The output of symbolic differentiation is the math expression of the derivative. On the other hand, automatic differentiation outputs numerical values of derivatives. The main idea behind automatic differentiation is that every numerical computation can be decomposed to a set of basic operations for which the derivatives are known. By applying the chain rule to these derivatives we can compute the value of the derivative of the overall expression.

Despite the difference in the underlying mechanics and the output, the programming interface to the user is pretty much the same in both approaches. The user calls a command or method to which it gives as input a numerical function. By combining techniques for computation of derivatives with mathematical optimization algorithms, systems for declarative machine learning can lead the way to automatic solution of a large class of objective functions.

\section{Optimization Techniques}
\label{sec:opt_techniques}
Optimization is an important part of the declarative paradigm. Since the user defines the result and not how this is computed, the system can provide a number of possible implementations and choose one that is good enough, if not optimal, in terms of efficiency. Considering the systems presented in the previous section, libraries of algorithms are used as black boxes and there is not much to be done on that front. Nevertheless, the other categories use either rule or cost-based approaches and borrow their ideas mainly from two areas: database systems and compilers. We describe how these ideas are applied on the declarative data science paradigm below.

The purpose of this section is to cover optimizations that transform the structure or the implementation of a program. Parallelization, scheduling and low-level optimizations regarding write operations, unless specifically relevant to the nature of data science tasks, are out of scope. Also systems that leverage optimizers of existing languages, such as BUDS and MLog, which translate their programs to SimSQL \cite{Cai13} and Datalog respectively, are not covered in the sections about optimization in this survey.

\subsection{Rule-based optimization}
\label{sec: rule_optimizations}
\begin{figure}
\captionsetup{justification=centering}
\centering
\includegraphics[width=0.7\textwidth]{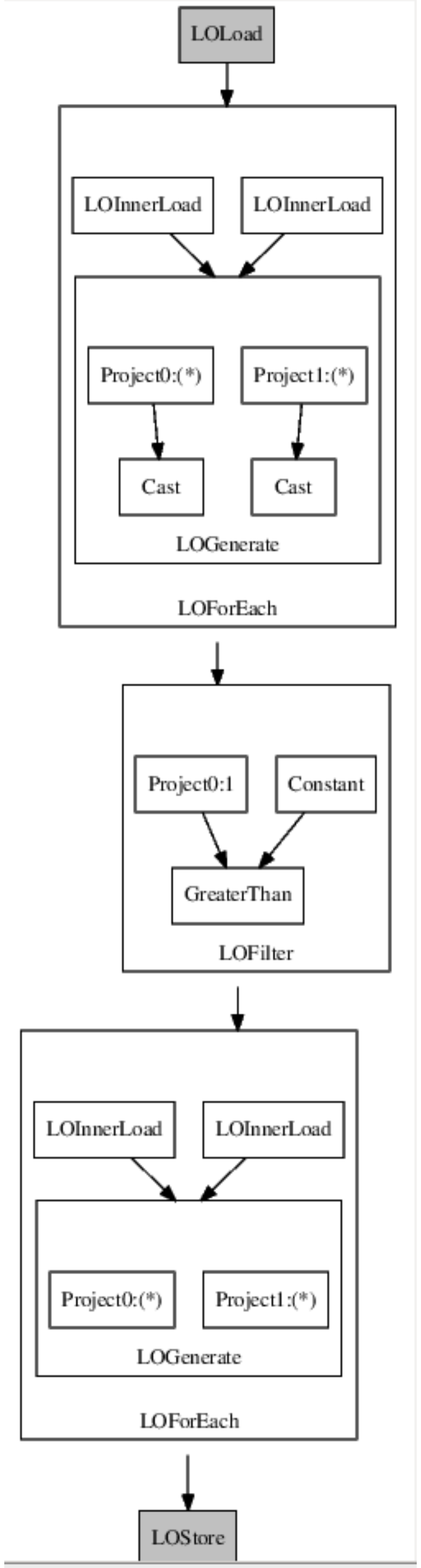}
\caption{Logical plan of the program in listing \ref{lst: pig_simple_program} generated by Pig}
\label{fig: pig_logical}
\end{figure}

This type of optimizations is based on rules. Every time certain conditions are met, these rules are triggered and perform specific transformations on the code. Languages that support relational operators, such as Pig Latin and Spark, can apply typical optimizations of database systems, e.g pushing filters closer to the data sources. In the following example in Pig Latin, we can notice that in the generated logical plan of figure \ref{fig: pig_logical} the filtering operation is performed before the {\ttfamily FOREACH} operator, whereas in the user code they are written in reversed order. Another optimization that takes place in the logical plan is the pruning of unused columns. The code states that four columns of the input data are loaded. However, as only the first two are used by the rest of the program, the optimizer transforms the logical plan so that only those two are eventually loaded. 

\begin{lstlisting}[caption={A simple program in Pig Latin}, label={lst: pig_simple_program}]
data = LOAD 'pig_test/weight_values.csv' USING PigStorage(',') AS (d0:double, d1:double, d2:double, d3:double);

data_projected = FOREACH data GENERATE $0, $1;

data_filtered = FILTER data_projected BY ($1)>5;

DUMP data_filtered;

STORE data_filtered INTO 'pig_test/data_filtered' USING PigStorage(',');
\end{lstlisting}

Other optimizations include pushing {\ttfamily LIMIT} operators near data or similarly for {\ttfamily FOREACH} operators when they are combined with {\ttfamily FLATTEN} operators. The main idea of these techniques is to decrease the number of tuples to be processed by expensive operations, such as join.  So, as {\ttfamily FLATTEN} unnests data and may increase the number of generated tuples, it is advised to be moved after the {\ttfamily FOREACH} operation. Finally, pipelining operations on data is also applied when it's possible. For example two consecutive {\ttfamily FOREACH} statements can be merged, in order to avoid a second pass on data. The capability to pipeline operations and avoid materialization of intermediate results is due to the lazy evaluation of programs that is supported by many of the described systems, such as Pig Latin, Jaql and Spark. This means that logical plans represent just a sequence of steps and are not physically executed until an output operation is requested, e.g. a print or store command, allowing the program to be optimized as a whole. More sophisticated optimizations, such as reordering of joins, are not well achieved with rules and depend on the computation of some cost metric, which is the topic of the next section.

Rule-based rewrites are also used by systems that provide linear algebra operators. For example, SystemML generates High-level Operator (HOP) plans that represent the data flow mainly between linear algebra operators, such as cell-wise multiplication, matrix multiplication or matrix transpose. Those plans are similar to the logical plans used in databases. An example of  a HOP DAG, which corresponds to the code snippet in listing \ref{lst: dml_code} is displayed in listing \ref{lst: dml_HOP_DAG}.

%\begin{lstlisting}[caption={Code sample in DML}, label={lst: dml_code}]
%computeGradientDescent = function (matrix[double] X,matrix[double] y,matrix[double] theta,
%double alpha) return (matrix[double] J,matrix[double] theta){
%	...
%	     m = nrow(y);
%		pred = X%*%theta;
%		d = (t(X)%*%(pred-y))/m;
%	...
%}
%\end{lstlisting}

\begin{lstlisting}[caption={Code sample in DML}, label={lst: dml_code}]
gradients = (t(features)%*%error)/m;
weights = weights - learning_rate*rowSums(gradients);
predictions = features%*%weights;
\end{lstlisting}
At the moment, SystemML does not provide a visualization tool for plans, but we can analyse this representation by identifying a few key components. The first number of each line is the HOP id and right next to it there is the operation code. Operation codes are divided into six categories: binary, unary, aggregate unary, aggregate binary, reorganize and data. So, b(-) is a binary operator that performs a subtraction between two cells of a matrix, or ba(+*) is an aggregate binary operator that first multiplies cells of a matrix and then sums the individual results. Numbers in parenthesis explain data dependencies between HOPs. For example, HOP 63 depends on the output of HOP 55. The first pair of brackets include characteristics of the output matrix, e.g. number of rows and columns, whereas the second pair of brackets provide memory estimates for input data, intermediate results and output matrices. Finally, CP (Single Node Control Program), SP (Spark) and MR (MapReduce) denote the execution type.

%\begin{lstlisting}[caption={HOP plan generated by SystemML for code in listing \ref{lst: dml_code}}]
%(2) TRead y [300,1,1000,1000,-1] [0,0,0 -> 0MB], CP
%(3) u(nrow) (2) [0,0,-1,-1,-1] [0,0,0 -> 0MB]
%(4) TWrite m (3) [0,0,0,0,-1] [0,0,0 -> 0MB], CP
%FOR (lines 21-29) [in-place=[J, d, pred]]
%GENERIC (lines 22-28) [recompile=false]	
%(22) TRead theta [14,1,1000,1000,-1] [0,0,0 -> 0MB], CP
%(16) TRead alpha [0,0,-1,-1,-1] [0,0,0 -> 0MB], CP
%(17) TRead X [300,14,1000,1000,-1] [0,0,0 -> 0MB], CP
%(24) r(t) (17) [14,300,1000,1000,-1] [0,0,0 -> 0MB]
%(23) ba(+*) (17,22) [300,1,1000,1000,-1] [0,0,0 -> 0MB]
%(18) TRead y [300,1,1000,1000,-1] [0,0,0 -> 0MB], CP
%(25) b(-) (23,18) [300,1,1000,1000,-1] [0,0,0 -> 0MB]
%(26) ba(+*) (24,25) [14,1,1000,1000,-1] [0,0,0 -> 0MB], CP
%(21) TRead m [0,0,0,0,-1] [0,0,0 -> 0MB], CP
%(27) b(/) (26,21) [14,1,1000,1000,-1] [0,0,0 -> 0MB], CP
%\end{lstlisting}

\begin{lstlisting}[caption={HOP plan generated by SystemML for code in listing \ref{lst: dml_code}}, label={lst: dml_HOP_DAG}]
(59) TRead weights [13,1,-1,-1,-1] [0,0,0 -> 0MB], CP
(55) TRead features [506,13,-1,-1,-1] [0,0,0 -> 0MB], CP
(63) r(t) (55) [13,506,-1,-1,-1] [0,0,0 -> 0MB]
(58) TRead error [506,1,-1,-1,-1] [0,0,0 -> 0MB], CP
(64) ba(+*) (63,58) [13,1,-1,-1,-1] [0,0,0 -> 0MB], CP
(66) b(/) (64) [13,1,-1,-1,-1] [0,0,0 -> 0MB], CP
(122) t(-*) (59,66) [13,1,-1,-1,-1] [0,0,0 -> 0MB], CP
(71) TWrite weights (122) [13,1,-1,-1,-1] [0,0,0 -> 0MB], CP
(72) ba(+*) (55,122) [506,1,-1,-1,-1] [0,0,0 -> 0MB], CP
\end{lstlisting}

This particular HOP plan begins with reading the weights and features that will be used to compute gradients. The computation of gradients starts at HOP 63, which outputs the transpose of matrix {\ttfamily features}. The result is read by the operator with id 64, which multiplies it with the {\ttfamily error} vector. The next operator divides the product with scalar {\ttfamily m}. HOP 122 updates the values of weights by subtracting the previously computed gradients from the current weight values. Finally, HOP 72 computes the product between {\ttfamily features} and {\ttfamily weights} in order to generate new predictions. The size of the matrices is recorded in the first pair of brackets. For example, features is a 506x13 matrix, but after applying a transpose operator it becomes a 13x506 matrix. It is important to note that HOPs are logical operators, placeholders in a sense, that will be replaced by specific implementations during the Low-level Operator (LOP) planning phase.

SystemML begins optimizing a HOP DAG by applying a number of static rewrites, i.e. cost-independent rewrites, which include common subexpression elimination, algebraic simplifications and removal of branches. Common subexpression elimination aims at computing subexpressions that are repeated inside code once, e.g.  if {\ttfamily (predictions-labels)} of listing \ref{lst:systemml} existed in more that one places in the code above, then all of its redundant appearances would be removed. TensorFlow also performs a common subexpression elimination step on the computation graph by removing repeated computations and accumulating them to a single node of the graph, which gets connected with every other node that needs this computation. 

Algebraic simplifications regard rewrites that are always beneficial no matter the dimensions of the matrices/vectors involved. For example, operations with one or zero, i.e $X/1$ or $X-0$, leave matrices unchanged and can be replaced by the matrix itself ($X$). Another example is the use of binary operations that can in fact be transformed to unary operations, such as X+X which can be rewritten to $2*X$. Finally, if-else blocks that depend on constant conditions are replaced with the body of the corresponding branch. This rewrite makes propagation of unconditional matrix sizes easier, which is important for applying cost-based optimization techniques in a latter phase. Mahout Samsara's optimizer applies similar rule-based rewritings on the logical plan, such as replacing binary with unary operators when both operands are the same matrix. The choice of physical operators is also based on heuristics regarding partitioning and key values, as well as types of distributed row matrices. However, overall Samsara provides a less advanced optimizer than SystemML, without support for cost-based optimization. In addition to this, the optimization methods that are used are much less documented and inspection of the source code seems to be the only available source of information at the moment.

In the intersection of machine learning and database systems, a novel research area that factorizes linear algebra computations into a series of relational operators has emerged \cite{Kumar2015}, \cite{Schleich2016}, \cite{AboKhamis2018}. These rewrites aim at reducing computational redundancy caused by joins. For example, when working on relational data, features are usually split across different tables, let us say $S$ and $R$. Hence, the product $w^{T}x$ between features and weights can be decomposed to inner products over the base tables $S$ and $R$. This decomposition reduces the redundancy that is caused by computing the inner product when the same tuple from table $R$ is joined with multiple tuples from table $S$. The partial inner products from $R$ can be stored in a relation and reused for every joined tuple in $S$.  Automating such rewrites remains a challenge, though. Recent work \cite{Chen2017} proposes heuristic rules for automatically translating a set of linear algebra operators, which are common in machine learning, into operations over normalized data. These rules are based on two simple metrics, whose role is important on the speedups yielded by the rewritings: tuple ratio (\#tuples of table1/\#tuples of table2) and feature ratio (\#columns of table1/\#columns of table2) of two joined tables. Whenever tuple ratio and feature ratio are below an experimentally tuned threshold, rewriting rules are not fired. Thresholds are conservative in the sense that they may wrongly disallow rewritings that could improve runtime.

%For example, in a linear regression implementation operating on normalized data and using gradient descent, one can decouple the parameter convergence from the costly multiplication between the feature table and its transpose, which is necessary in every iteration to compute $ \sum_{i=1}^{n}(\sum_{j=0}^m\theta_{j}x_{i,j}-y_{i})x_{i,j}$, albeit remains unchanged between iterations. This step can be pre-computed once, stored in a relation and used in every iteration afterwards. To do so, \cite{Schleich2016} proposes to rewrite the expression above so that label $y_{i}$ becomes part of the product between $\theta$ and $x$ by adding a parameter $\theta=-1$. Similarly   
Apart from the rule-based rewritings described above, there are also a few simple rules that come from the domain of compilers. These include techniques like function and variable inlining. Jaql, SystemML and Tupleware already exploit ideas from this area. Function inlining replaces a function call with the code of the function, whereas parameters become local variables. Variable inlining is similar, so when a variable is used in an expression, it is replaced by its definition, which is either an expression or a value. Inlining reduces the overhead caused by functions calls and variable references, but at the same time increases memory cost and is usually avoided for large functions.

\begin{figure}
\captionsetup{justification=centering}
\centering
\includegraphics[width=\textwidth]{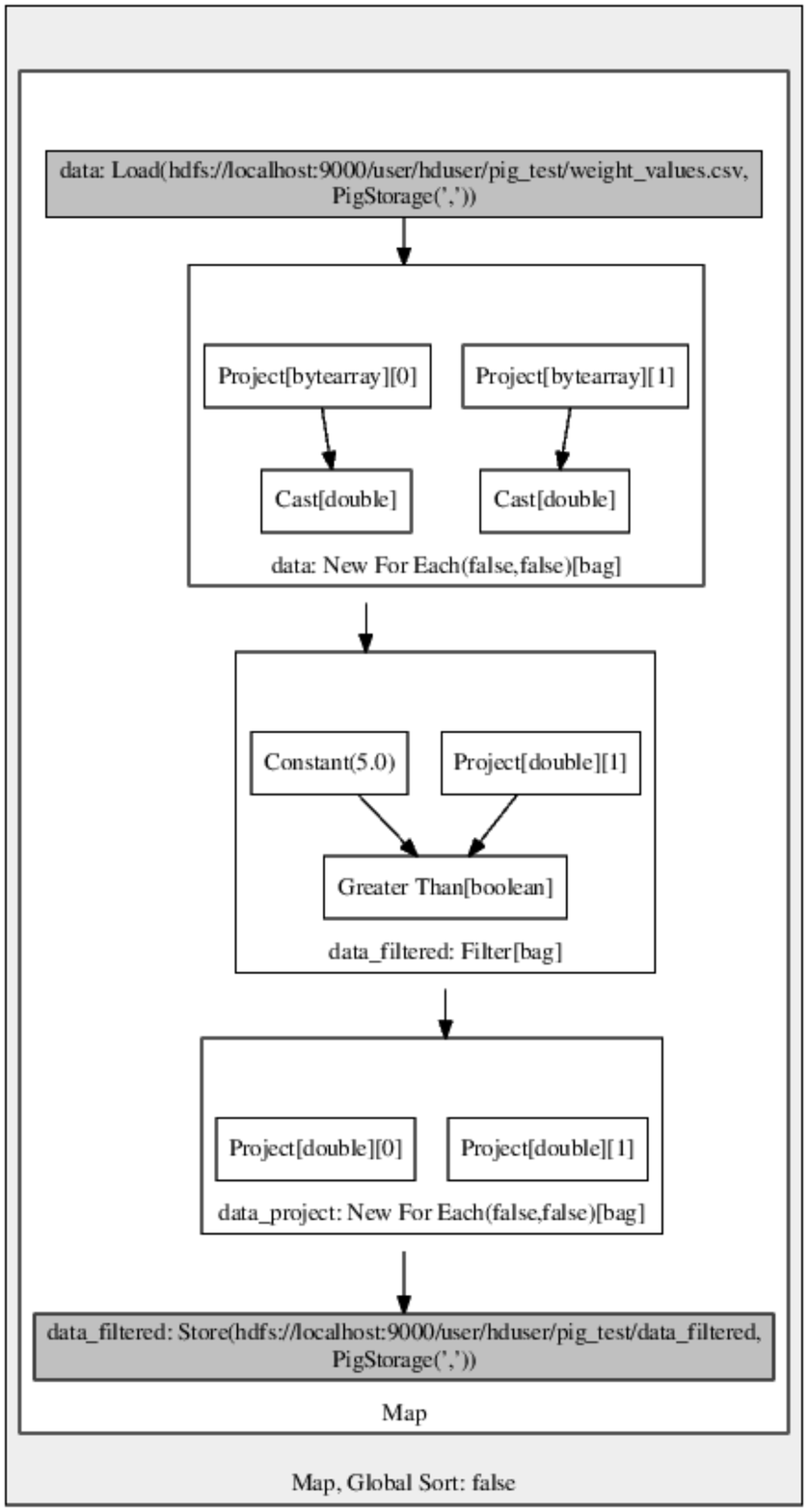}
\caption{MapReduce plan of the program in listing \ref{lst: pig_simple_program} generated by Pig}
\label{fig: pig_mapreduce}
\end{figure}

Finally, many of the systems for large-scale analytics need rules to translate programs written in higher-level languages into the programming model of the preferred execution engine efficiently. For example, Jaql and Pig that translate their programs to the MapReduce model, need to create as few MapReduce jobs as possible in order to avoid materialization of data between jobs. To do so, they identify steps of the logical plan which are mappable, i.e. can be executed independently over partitions of data (for example {\ttfamily FILTER} and {\ttfamily FOREACH} operators), and group them into a single Map function. Expressions that are encountered after a group by operator are executed inside a Reduce function. They also employ ways to translate group operators with multiple inputs, such as {\ttfamily COGROUP}, either by creating separate map functions for each input and aggregating them in a single reduce, or by adding a field which registers each tuple to the dataset it belongs to. The MapReduce plan for the Pig Latin program in listing \ref{lst: pig_simple_program} is displayed in figure \ref{fig: pig_mapreduce}. We can see that all operations are grouped inside a single map function, as Pig takes advantage of the fact that {\ttfamily FILTER} and {\ttfamily FOREACH} operations can be run in parallel over partitions of data. In a similar way of thinking, SciDB's optimization at the physical level focuses on minimization of data movement and efficient parallelization. It follows an adaptive approach by identifying subtrees of the physical plan where operators can be pipelined and as a consequence parallelized over a cluster of machines. For the rest of the plan that cannot be parallelized, a scatter/gather operator is used, which gathers the local chunks of the nodes in a buffer and pushes it to the node where the data need to be transferred. Such optimizations depend highly on the properties of the execution engine that each system uses. For example, if we switch from Hadoop to Spark we can avoid many of the drawbacks that come with materialization of data between jobs.

\subsection{Cost-based optimization}
\label{sec: cost_optimizations}
Another type of optimization techniques is based on the computation of cost metrics. That means that different rewritings of a plan are evaluated using a cost model and it depends on the evaluation which of the rewritings will take place. On the contrary, rule-based optimization always triggers a rule, when conditions apply, even if the rewriting does not improve the generated plan. Systems, such as Tupleware and Spark, support cost-based optimization on relational operators either at logical or at physical level, using well-studied ideas from database optimization. The cost model for relational operators is based on statistics of tables that are collected periodically, e.g. number of tuples, selectivity of columns. For example, Tupleware performs join reordering based on data statistics, whereas Spark's Catalyst optimizer uses only ruled-based optimization at the logical level, but chooses implementations for join operators at the physical level by estimating I/O operations.

A more novel area of cost-based optimization addresses linear algebra operations. SystemML seems to be the most advanced system in this direction. Cost-dependent rewrites are applied on expensive operations of the logical plan, i.e. HOP DAG, and the cost is based on a number of metrics, including matrix dimensions, floating point operations, I/O operations and shuffle cost, depending on each optimization phase. Some of the first rewritings of a DAG consisting of linear algebra operators include the removal of operators when one of the matrices contains only zeros, the replacement of row and column sums with table sums whenever the matrix consists of a single row or column, or the removal of indexing whenever the dimensions of the matrices are the same. These transformations are size-dependent, as the optimizer needs to be aware of the dimensions and the sparsity of the input matrices in order to perform them. That being said, there is still no cost function that is minimized / maximized on this phase. We can observe some of these rewritings in the HOP plan of the code snippet in listing \ref{lst: dml_code_2}.

\begin{lstlisting}[caption={Code sample in DML}, label={lst: dml_code_2}]
gradients = (t(features)%*%error)/m;
weights = weights - learning_rate*rowSums(gradients);
\end{lstlisting}
The generated plan for this part is in listing \ref{lst: hop_plan_2}.
\begin{lstlisting}[caption={HOP plan generated by SystemML for code in listing \ref{lst: dml_code_2}}, label={lst: hop_plan_2}]
(59) TRead weights [13,1,-1,-1,-1] [0,0,0 -> 0MB], CP
(55) TRead features [506,13,-1,-1,-1] [0,0,0 -> 0MB], CP
(63) r(t) (55) [13,506,-1,-1,-1] [0,0,0 -> 0MB]
(58) TRead error [506,1,-1,-1,-1] [0,0,0 -> 0MB], CP
(64) ba(+*) (63,58) [13,1,-1,-1,-1] [0,0,0 -> 0MB], CP
(66) b(/) (64) [13,1,-1,-1,-1] [0,0,0 -> 0MB], CP
(122) t(-*) (59,66) [13,1,-1,-1,-1] [0,0,0 -> 0MB], CP
\end{lstlisting}
The plan begins with fetching training data and weight values and continues with transposing matrix {\ttfamily features} as well as performing the matrix multiplication with the already computed {\ttfamily error} matrix. After the computation of vector {\ttfamily gradients} at HOP 66, one can observe that the unary aggregation of row sums of {\ttfamily gradients} is removed, as the number of columns is equal to one and this makes it the same with multiplying with the single cell of each row. So, the expression 
\begin{lstlisting}[label={lst: dml_code_3}]
weights = weights - learning\_rate*rowSums(gradients)
\end{lstlisting}
is transformed to just
\begin{lstlisting}[label={lst: dml_code_4}]
weights = weights-learning\_rate*gradients
\end{lstlisting}

Another example is the expression in listing \ref{lst: dml_code_5}.
\begin{lstlisting}[caption={Code sample in DML}, label={lst: dml_code_5}]
error = colSums(features%*%weights - labels)/nrow(features);
\end{lstlisting}
We can observe in the HOP plan of listing \ref{lst: hop_plan}, that the optimizer has pushed a summation over columns of {\ttfamily features} (HOP id 128) before computing the product with {\ttfamily weights}, in order to reduce the size of the multiplication. 

\begin{lstlisting}[caption={HOP plan generated by SystemML for code in listing \ref{lst: dml_code_5}}, label={lst: hop_plan}]
(92) TRead features [506,13,-1,-1,-1] [0,0,0 -> 0MB], CP
(128) ua(+C) (92) [1,13,-1,-1,-1] [0,0,0 -> 0MB], CP
(93) TRead weights [13,1,-1,-1,-1] [0,0,0 -> 0MB], CP
(133) ba(+*) (128,93) [1,1,-1,-1,-1] [0,0,0 -> 0MB], CP
(134) u(cast_as_scalar) (133) [0,0,0,0,-1] [0,0,0 -> 0MB]
(94) TRead labels [506,1,-1,-1,-1] [0,0,0 -> 0MB], CP
(126) ua(+RC) (94) [-1,-1,-1,-1,-1] [0,0,0 -> 0MB], CP
(127) b(-) (134,126) [0,0,-1,-1,-1] [0,0,0 -> 0MB], CP
(124) u(cast_as_matrix) (127) [1,1,1000,1000,-1] [0,0,0 -> 0MB]
(103) b(/) (124) [1,1,-1,-1,-1] [0,0,0 -> 0MB], CP
\end{lstlisting}
SystemML's optimizer examines the dimensions of the matrices involved and notes that the initial code will multiply two matrices of size (506x1)*(13x1), whereas pushing the summation first will result in a multiplication of size (1x13)*(13x1). The same optimization is also applied on {\ttfamily labels} by fully aggregating over both rows and columns (HOP id 126), which results in executing a subtraction between scalars instead of matrices. After these two rewritings, the initial expression is finally transformed to

\begin{lstlisting}[label={lst: dml_code_6}]
error = ((colSums(features)\%*\%weights) - sum(label))/nrow(features)
\end{lstlisting}

SystemML's optimizer is able to propagate matrix sizes to the whole program, using a bottom-up procedural analysis inside and across DAGs. Starting from read operators, for which input sizes can be inferred or are known due to metadata, dimensions and sparsity properties are propagated to the operators of a DAG and finally to its result variables. Based on the semantics of linear algebra operators, computing the output matrix size of an operator is possible. In case of if or while conditions, the propagation procedure takes into account whether variable sizes change inside the loop and need to be re-propagated or whether both if and else conditions will result to the same output size.

Although the rewritings presented above are cost-based, there is no search algorithm that evaluates different plans in order to find the optimal one. However, when it comes to matrix multiplication chains, SystemML uses dynamic programming in order to find the multiplication order that minimizes the number of operations needed to compute the final product.

The rewritings described above can be implemented in any system that supports linear algebra operations. In general though, linear algebra operators are not widely commutable and do not offer many opportunities for reorderings at the logical level, as it applies to relational operators. Apart from altering the order of matrix multiplication chains, most of the rewritings at the logical level are based on simple mathematical properties. When linear algebra operators are integrated to a relational database as suggested in \cite{Luo17}, reorderings between relational and linear algebra operators can also be applied based on the dimensions of the matrices. For example, in a SQL query that involves both joining tables and multiplying matrices, the order of joins can play an important role on the size of the result of matrix multiplication.

Cost-based optimization techniques can also be applied on physical plans that consist of linear algebra operators. The most advanced system at this front is again SystemML, which chooses different implementations for linear algebra operators based on cost functions. Its physical plans, LOP DAGs, consist of the physical operators for each high-level (logical) operator of the HOP DAG. The LOP DAG that corresponds to the HOP DAG of listing \ref{lst: hop_plan_2} is displayed in listing \ref{lst:lop_dag} . There may be implementations of LOPs for various runtime engines (single node, MapReduce and Spark), which is indicated by the first token of each runtime instruction in the LOP DAG, and more than one implementations of an operator for a specific execution engine. 

The choice between single-node and distributed execution is based on memory estimates for each HOP and the available budget of a single machine. Memory estimates assume single-threaded execution and are computed recursively from the leafs of a HOP DAG to its internal nodes using a precise model of dense and sparse matrices. The memory for internal HOPS is the sum of the estimates from their child nodes, the memory for intermediate results and the output memory of the HOP. Hence, HOPS that need less memory than the available budget in a single machine, will be executed locally, since local in-memory execution is assumed to be always cheaper than distributing data over the network. In case CP mode is not possible, the optimizer chooses a distributed LOP implementation based on a different cost function that takes into account I/O cost, shuffle cost and the degree of parallelism. Shuffle cost is the cost to redistribute data from mappers to the appropriate reducers on a distributed framework. It involves the time to write and read data from the file system divided by the number of mappers and reducers. For example, in case of a multiplication between a matrix and a small vector, the optimizer can choose to send a copy of the vector in every machine and perform local partial aggregations of the results, in order to avoid shuffle costs.

In the LOP plan of listing \ref{lst:lop_dag}, we can see that some operators are executed on a single node, whereas others are executed on Spark. This is because memory estimates for these operators surpass the available memory on the local machine, i.e. the available memory budget locally was 693MB, whereas the required memory for transposing matrix {\ttfamily X\_train} was 1990MB. We can also observe that different physical implementations are provided for the same HOP, e.g. cpmm LOP in line 5, which is a cross product based algorithm for distributed matrix multiplication.

\begin{lstlisting}[caption={Sample LOP plan corresponding to listing \ref{lst: hop_plan_2}}, label={lst:lop_dag}]
CP createvar _mVar22 scratch_space//_p3813_127.0.1.1//_t0/temp18 true MATRIX binaryblock -1 -1 1000 1000 -1 copy
SPARK r' X_train.MATRIX.DOUBLE _mVar22.MATRIX.DOUBLE
CP * 2.SCALAR.INT.true 506.SCALAR.INT.true _Var23.SCALAR.INT
CP createvar _mVar24 scratch_space//_p3813_127.0.1.1//_t0/temp19 true MATRIX binaryblock -1 1 1000 1000 -1 copy
SPARK cpmm _mVar22.MATRIX.DOUBLE red.MATRIX.DOUBLE _mVar24.MATRIX.DOUBLE MULTI_BLOCK
CP rmvar _mVar22
CP createvar _mVar25 scratch_space//_p3813_127.0.1.1//_t0/temp20 true MATRIX binaryblock -1 1 1000 1000 -1 copy
SPARK / _mVar24.MATRIX.DOUBLE 506.SCALAR.INT.true _mVar25.MATRIX.DOUBLE
CP rmvar _mVar24
CP createvar _mVar26 scratch_space//_p3813_127.0.1.1//_t0/temp21 true MATRIX binaryblock -1 1 1000 1000 -1 copy
\end{lstlisting}

%\begin{lstlisting}[caption={LOP plan corresponding to HOP plan in listing \ref{lst: hop_plan}}]
%CP createvar _mVar17 scratch_space//_p9561_127.0.1.1//_t0/temp14 true MATRIX binaryblock 1 506 1000 1000 -1 copy
%CP r' red.MATRIX.DOUBLE _mVar17.MATRIX.DOUBLE 1
%CP createvar _mVar18 scratch_space//_p9561_127.0.1.1//_t0/temp15 true MATRIX binaryblock 1 14 1000 1000 -1 copy
%CP ba+* _mVar17.MATRIX.DOUBLE X_train.MATRIX.DOUBLE _mVar18.MATRIX.DOUBLE 1
%CP rmvar _mVar17
%CP createvar _mVar19 scratch_space//_p9561_127.0.1.1//_t0/temp16 true MATRIX binaryblock 14 1 1000 1000 -1 copy
%CP r' _mVar18.MATRIX.DOUBLE _mVar19.MATRIX.DOUBLE 1
%CP rmvar _mVar18
%CP createvar _mVar20 scratch_space//_p9561_127.0.1.1//_t0/temp17 true MATRIX binaryblock 14 1 1000 1000 -1 copy
%CP / _mVar19.MATRIX.DOUBLE 506.SCALAR.INT.true _mVar20.MATRIX.DOUBLE
%CP rmvar _mVar19
%CP createvar _mVar21 scratch_space//_p9561_127.0.1.1//_t0/temp18 true MATRIX binaryblock 14 1 1000 1000 -1 copy
%CP -* theta.MATRIX.DOUBLE 0.09.SCALAR.DOUBLE.true _mVar20.MATRIX.DOUBLE _mVar21.MATRIX.DOUBLE
%\end{lstlisting}

Finally, in the context of work for in-database linear algebra \cite{Luo17} described in section \ref{sec: indb_lr}, preliminary ideas on executing plans which involve both relational and linear algebra operators more efficiently are explored. For example, when combining matrix multiplication with join operators,  an optimizer aware of the dimensions of the involved matrices can choose a better plan. It could choose to perform matrix multiplication before joining, in order to reduce the size of the matrices moving up the plan, instead of performing a series a joins and leave matrix multiplication for the end. 

\subsection{Optimization and UDFs}
\label{sec: udf_optimizations}
Systems that adhere to the class of extensions of the MapReduce model depend heavily on user-defined functions, as their operators are second-order functions. These UDFs are written in functional or imperative languages, such as Scala or Java, and the semantics of the code is unknown. As a result, optimization of UDFs is more difficult and the techniques that are currently applied by such systems are quite limited. In this section, we describe methods that tackle some of the UDF optimization issues and were proposed by Flink and Tupleware. Spark, despite making heavy use of second-order functions, does not provide optimization techniques for their content. 

Flink uses static code analysis to determine which data are read and written from each operator and separate them into read and write sets. By checking whether read and write sets overlap, we are able to know whether a reordering between two operators would result to a semantically equivalent plan. This idea is quite similar to the concurrency control techniques used in databases. So when only the read sets of two operators overlap, reordering of them will not break the semantics of the program. On the other hand, when read and write sets overlap, we are not able to ensure semantic equivalence. In case of group by operators, the optimizer can also determine whether the cardinality of the grouping attribute will remain unchanged after a reordering. That ensures that the input size of the grouping operator will also be the same. It is clear that this method still does not understand the semantics of the code and as a consequence it is conservative in the sense that it probably forbids valid reorderings and misses optimization opportunities.

Apart from the logical phase, Flink exploits read and write sets of UDFs in optimization of physical plans. By knowing whether a UDF modifies a partitioning or sorting key, it is possible to determine whether physical data properties change. For example, if a grouping operator has already partitioned data based on the same key that a join operator will be applied on, there is no need to reshuffle data and increase network I/O. However, this technique is not enough to provide an accurate I/O cost for each plan and Flink at the moment bases its cost estimations on hints for UDF selectivity provided by the user or derived from earlier phases.

Tupleware also introspects UDFs by examining their LLVM intermediate representation, but focuses on other types of information. The purpose of its analysis is to determine vectorizability and estimate CPU and memory requirements of the UDF code. Vectorizability achieves data level parallelism by using the registers of a processor to apply a single instruction on multiple data simultaneously. CPU cycles can provide an estimation of compute time, whereas memory bandwidth can be used to predict load time of operands. Therefore, these two metrics can estimate whether a problem is compute-bound or memory-bound by comparing compute to load time. These statistics allow Tupleware to employ an adaptive strategy that switches between pipelining and bulk processing. 

As an example consider map operators. It's common to group consecutive maps to a single pipeline, in order to leverage data locality. In addition to this, Tupleware can identify which of the map UDFs are vectorizable, exploit SIMD vectorization for them and cache intermediate results to avoid delays. In case there is one or more vectorizable UDFs at the beginning of the pipeline, Tupleware's optimizer can examine statistics of load time to evaluate whether fetching UDF operands would be faster than UDF computations. Based on this evaluation, it will then determine whether it should apply SIMD vectorization or stick to the initial operator pipelining.

Concerning reduce operators, the construction of the hash table can also be parallelized using SIMD vectorization. Moreover, in case reduce is based on a single key and the aggregation function is commutative and associative, Tupleware again computes partial aggregates in parallel using vectorization and combines them at the end to derive the final result.

UDFs in imperative languages can also be combined with SQL code. Their use is not encouraged though \cite{t-sql-good-bad}, \cite{functions-wreck-performance}, \cite{performance-considerations} due to the performance overhead that comes with their evaluation strategy and the fact that they are not processed by the query optimizer. In an effort to address these issues, there has been work on optimization of UDFs used in SQL queries \cite{Simhadri14}, \cite{Ramachandra17}. The main approach of this line of work is to translate imperative code in UDFs to SQL subqueries, which then substitute UDF expressions in the query tree. The execution of subqueries is performed using the \textit{Apply} operator. The supported imperative constructs include variable declarations and assignments, branching, return statements and cursor loops.        

\section{Comparison and Discussion}
\label{sec:comparison}

After presenting a broad range of systems for data analytics, we summarize how the described systems align with the properties of Section \ref{sec:scope}. TileDB and Model Selection Management Systems of section \ref{sec:msms} are excluded. The former currently provides only storage management. For the latter we explain why the properties are not applicable in section \ref{sec:msms}. Table \ref{tab:2} summarizes the results of this analysis. The last column of the table concerns the scope of the systems with regard to the following data science areas: data processing (DP - relational operators and other data transformations), machine learning (ML - linear algebra operators), convex optimization (LP/QP - linear/quadratic programming). The term "machine learning" refers to operators for building machine learning algorithms, not black box machine learning libraries. Table \ref{tab:3} also explains the types of plan optimizations that are provided by each system.

\begin{longtable}{p{1.5cm}|p{1.8cm}|p{0.6cm}|p{0.6cm}|p{1.2cm}|p{1.2cm}|p{1.4cm}|p{1.5cm}|p{0.7cm}}
%\centering
\caption{Declarativity of systems based on seven properties}\\
\label{tab:2}       % Give a unique label
%\begin{tabular}{|m{1.2cm}|m{1.2cm}|m{1.2cm}|m{1.2cm}|m{1.2cm}|m{1.2cm}|m{1.2cm}|m{1.2cm}|m{1.2cm}|}
%\hline
System/Language & Independence of Data Abstractions & DP Ops & ML Ops & Plan Optimization & Lack of Control Flow & Automatic Computation of Solution & Lack of Code with Unknown Semantics & Scope\\
\hline
Pig Latin & \checkmark & \checkmark & & \checkmark & \checkmark & & \checkmark & DP \\
\hline
Jaql & \checkmark & \checkmark & & \checkmark & \checkmark & & \checkmark & DP \\ 
\hline
U-SQL & \checkmark & \checkmark & & \checkmark & \checkmark & & \checkmark & DP\\
\hline
Spark & & \checkmark & \checkmark & \checkmark & & & & DP, ML \\
\hline
Flink (Stratosphere) &  & \checkmark &  & \checkmark & & & & DP\\
\hline
DryadLINQ &  & \checkmark & & \checkmark & & & & DP\\
\hline
Tupleware & & \checkmark & & \checkmark & & & & DP \\
\hline
SystemML & \checkmark & & \checkmark & \checkmark & & & \checkmark & ML \\
\hline
Mahout Samsara & & & \checkmark & \checkmark & & & \checkmark & ML \\
\hline
BUDS & \checkmark & & \checkmark & & \checkmark & & \checkmark & ML \\
\hline
TensorFlow & & & \checkmark & \checkmark & & \checkmark & \checkmark & ML \\
\hline
SciDB & \checkmark & \checkmark & \checkmark & \checkmark & \checkmark & & \checkmark & DP, ML \\
\hline
LogicBlox & \checkmark & \checkmark & & \checkmark & \checkmark & \checkmark & \checkmark & DP, LP/QP\\
\hline
MLog & & & \checkmark & & \checkmark & \checkmark & \checkmark & ML \\
\hline
ReLOOP & & \checkmark & & & & \checkmark & \checkmark & LP/QP \\
\hline
An extension of SQL with linear algebra \cite{Luo17} & \checkmark & \checkmark & \checkmark & \checkmark & \checkmark & & \checkmark & DP, ML\\
\hline
%\end{tabular}
\end{longtable}

Based on these properties, we can see from table \ref{tab:2} that the most declarative approaches include: SciDB, LogicBlox and the extension of SQL with linear algebra operators. These systems provide or emulate both data processing and commonly used operations in machine learning, offer plan optimization and the programs in the languages they support do not depend on a particular execution order. Moreover, user defined functions are only necessary when a task cannot be expressed using the primitives of the language, since operators in these languages do not serve as second-order functions. It is important to note that SystemML, despite not fully declarative according to the properties of the table, is a strong player in the area as an Apache project focused on large-scale machine learning with increasing community adoption and extensive available documentation, which makes new users' on-boarding easy.

\begin{table}
\centering
\caption{Optimization capabilities per system}
\label{tab:3}       % Give a unique label
\begin{tabular}{|m{1.5cm}|m{1.5cm}|m{1.5cm}|m{1.5cm}|m{1.5cm}|m{1.5cm}|m{1.5cm}|}
\hline
System or Language & Relational Algebra & Linear Algebra & User Defined Functions & Compiler-Based Optimizations & Common Sub-expression Elimination & Automatic Parallelization\\
\hline
Pig Latin & \checkmark & & & & & \checkmark \\
\hline
Jaql & \checkmark & & & \checkmark & & \checkmark \\ 
\hline
U-SQL & \checkmark & & & & & \checkmark \\
\hline
Spark & \checkmark & & & & & \\
\hline
Flink \/ Stratosphere & \checkmark & & \checkmark & & & \\
\hline
DryadLINQ & & & & & \checkmark & \checkmark \\
\hline
Tupleware & \checkmark & & \checkmark & \checkmark & & \\
\hline
SystemML & & \checkmark & & & \checkmark & \checkmark \\
\hline
Mahout Samsara & & \checkmark & & & & \checkmark \\
\hline
TensorFlow & & & & & \checkmark & \\
\hline
SciDB & & \checkmark & & & & \checkmark\\
\hline
LogicBlox & \checkmark & & & & & \\
\hline
An extension of SQL with linear algebra \cite{Luo17} & \checkmark & \checkmark & & & & \\
\hline
\end{tabular}
\end{table}

Regarding most declarative systems, there are still differences in their design choices. SciDB supports an array-based data model, instead of a relational one. LogicBlox and the extension of SQL with linear algebra operators both work directly on relational data. The LogicBlox database follows the model+solver paradigm, where models are defined as Datalog programs and the solution is provided by convex optimization solvers, whereas an extension of SQL would natively support linear algebra operators, but the mathematical optimization algorithm, e.g. gradient descent need also be coded by the user. 
%Transforming relational data to the array model may require considerable effort or most significantly may not be practical.

These two approaches emphasize the power of the relational model and the benefits of keeping data normalized. Despite the limitations of relational algebra when it comes to looping, it is clear that its operators and the concept of relation serve preprocessing and feature engineering tasks very well and therefore cannot be dropped from the machine learning toolkit. Based on these observations, two main directions are forming in the area of declarative machine learning. One direction claims that we should give up our obsession with the database paradigm and use different platforms for machine learning and data storage. Advocates of this direction seem to believe that the current situation will become the status quo and the idea of integrating machine learning into the database will eventually fade away due to usability, performance and expressivity challenges that would be not addressed in an efficient manner. In this case however, an open question remains: how close are we in achieving declarative specification of machine learning algorithms when we program them in systems that use imperative languages, where the execution order of commands matters? Notice that none of the systems in this category scores high in the seven properties of table \ref{tab:2}. The other direction estimates that history will repeat itself. As it happened with other processing needs in the past that were ultimately integrated with database architectures, such as stream processing or full text search, at some point it will come down to adding a suitable set of operators to the relational algebra or merging relational with linear algebra, in order to provide a unified environment for data science.Although the approaches that show high declarativity according to our framework set a good foundation for this direction, the problem of declarative data science and machine learning cannot be considered solved.

\section{Conclusion}
\label{sec:conclusion}
In this paper, we presented an extensive survey over systems in the area of declarative data analytics. As described in the previous sections, the area can be divided to a number of categories, each one following a different programming model and degree of declarativity. Today, data scientists use a hairball of these systems, as each of them fits best different scenarios. It is thus frequent to build pipelines of various frameworks to support all the components of a machine learning solution. Building such pipelines of course involves lots of glue code, as well as tedious ETL processes, and requires at least a basic level of familiarity with a number of systems and languages. As a result, those approaches are not easily maintained and require different levels and aspects of engineering expertise. We argue that the challenges described above and the solutions that will surround them, form the driving force that will determine which direction will prevail, database management systems or general-purpose machine learning systems operating on denormalized data. Based on our analysis, it seems that despite early trends in the area favouring Map-Reduce based frameworks, the database ecosystem strikes back and proposes promising approaches in terms of expressivity and efficiency, in order to provide a holistic framework for a wide range of analytic tasks, including machine learning, in the same environment where data lives. A database system with such capabilities would definitely be a great step to democratize data science and machine learning. 

\begin{acks}
We thank Panagiotis-Ioannis Betchavas for the implementation of Linear Regression using DML in section \ref{sec:ml_frameworks}.
\end{acks}

\bibliographystyle{ACM-Reference-Format}      % mathematics and physical sciences
\bibliography{../survey_bibliography}   % name your BibTeX data base

\end{document}